\documentclass[twocolumn]{aastex63}

\newcommand{\clean}{CLEAN}
\newcommand{\tclean}{\texttt{tclean}}

\received{}
\revised{}
\accepted{}
\submitjournal{AJ}

\shorttitle{MAPS I}
\shortauthors{\"Oberg et al.}
\graphicspath{{./}{}}

\begin{document}

\title{Molecules with ALMA at Planet-forming Scales (MAPS) I: Program Overview and Highlights.}

\correspondingauthor{Karin \"Oberg}
\email{koberg@cfa.harvard.edu}

\author[0000-0001-8798-1347]{Karin I. \"Oberg}
\affiliation{Center for Astrophysics \textbar\ Harvard \& Smithsonian, 60 Garden St., Cambridge, MA 02138, USA}

\author[0000-0003-4784-3040]{Viviana V. Guzm\'an}
\affil{Instituto de Astrof\'isica, Pontificia Universidad Cat\'olica de Chile, Av. Vicu\~na Mackenna 4860, 7820436 Macul, Santiago, Chile}

\author[0000-0001-6078-786X]{Catherine Walsh}
\affiliation{School of Physics and Astronomy, University of Leeds, Leeds, UK, LS2 9JT}

\author[0000-0003-3283-6884]{Yuri Aikawa}
\affiliation{Department of Astronomy, Graduate School of Science, The University of Tokyo, Tokyo 113-0033, Japan}

\author[0000-0003-4179-6394]{Edwin A. Bergin}
\affiliation{Department of Astronomy, University of Michigan, 323 West Hall, 1085 South University Avenue, Ann Arbor, MI 48109, USA}

\author[0000-0003-1413-1776]{Charles J. Law}\affiliation{Center for Astrophysics \textbar\ Harvard \& Smithsonian, 60 Garden St., Cambridge, MA 02138, USA}

\author[0000-0002-8932-1219]{Ryan A. Loomis}\affiliation{National Radio Astronomy Observatory, 520 Edgemont Rd., Charlottesville, VA 22903, USA}

\author[0000-0002-2692-7862]{Felipe Alarc\'on }
\affiliation{Department of Astronomy, University of Michigan, 323 West Hall, 1085 South University Avenue, Ann Arbor, MI 48109, USA}

\author[0000-0003-2253-2270]{Sean M. Andrews} \affiliation{Center for Astrophysics \textbar\ Harvard \& Smithsonian, 60 Garden St., Cambridge, MA 02138, USA}

\author[0000-0001-7258-770X]{Jaehan Bae}
\altaffiliation{NASA Hubble Fellowship Program Sagan Fellow}
\affil{Earth and Planets Laboratory, Carnegie Institution for Science, 5241 Broad Branch Road NW, Washington, DC 20015, USA}
 \affiliation{Department of Astronomy, University of Florida, Gainesville, FL 32611, USA}

\author[0000-0002-8716-0482]{Jennifer B. Bergner}
\altaffiliation{NASA Hubble Fellowship Program Sagan Fellow}
\affiliation{University of Chicago Department of the Geophysical Sciences, Chicago, IL 60637, USA}

\author[0000-0002-8692-8744]{Yann Boehler}\affiliation{Univ. Grenoble Alpes, CNRS, IPAG, F-38000 Grenoble, France}

\author[0000-0003-2014-2121]{Alice S. Booth} \affiliation{Leiden Observatory, Leiden University, 2300 RA Leiden, the Netherlands}
\affiliation{School of Physics and Astronomy, University of Leeds, Leeds, LS2 9JT, UK}

\author[0000-0003-4001-3589]{Arthur D. Bosman}
\affiliation{Department of Astronomy, University of Michigan, 323 West Hall, 1085 South University Avenue, Ann Arbor, MI 48109, USA}

\author[0000-0002-0150-0125]{Jenny K. Calahan} 
\affiliation{Department of Astronomy, University of Michigan, 323 West Hall, 1085 South University Avenue, Ann Arbor, MI 48109, USA}

\author[0000-0002-2700-9676]{Gianni Cataldi}
\affiliation{National Astronomical Observatory of Japan, 2-21-1 Osawa, Mitaka, Tokyo 181-8588, Japan}
\affiliation{Department of Astronomy, Graduate School of Science, The University of Tokyo, Tokyo 113-0033, Japan}

\author[0000-0003-2076-8001]{L. Ilsedore Cleeves}
\affil{Department of Astronomy, University of Virginia, Charlottesville, VA 22904, USA}

\author[0000-0002-1483-8811]{Ian Czekala}
\altaffiliation{NASA Hubble Fellowship Program Sagan Fellow}
\affiliation{Department of Astronomy and Astrophysics, 525 Davey Laboratory, The Pennsylvania State University, University Park, PA 16802, USA}
\affiliation{Center for Exoplanets and Habitable Worlds, 525 Davey Laboratory, The Pennsylvania State University, University Park, PA 16802, USA}
\affiliation{Center for Astrostatistics, 525 Davey Laboratory, The Pennsylvania State University, University Park, PA 16802, USA}
\affiliation{Institute for Computational \& Data Sciences, The Pennsylvania State University, University Park, PA 16802, USA}
\affiliation{Department of Astronomy, 501 Campbell Hall, University of California, Berkeley, CA 94720-3411, USA}

\author[0000-0002-2026-8157]{Kenji Furuya} \affiliation{National Astronomical Observatory of Japan, 2-21-1 Osawa, Mitaka, Tokyo 181-8588, Japan}

\author[0000-0001-6947-6072]{Jane Huang}
\altaffiliation{NASA Hubble Fellowship Program Sagan Fellow}
\affiliation{Department of Astronomy, University of Michigan, 323 West Hall, 1085 South University Avenue, Ann Arbor, MI 48109, USA}
\affiliation{Center for Astrophysics \textbar\ Harvard \& Smithsonian, 60 Garden St., Cambridge, MA 02138, USA}

\author[0000-0003-1008-1142]{John~D.~Ilee}
\affiliation{School of Physics and Astronomy, University of Leeds, Leeds, UK, LS2 9JT}

\author[0000-0002-2358-4796]{Nicolas T. Kurtovic}
\affiliation{Departamento de Astronom\'ia, Universidad de Chile, Camino El Observatorio 1515, Las Condes, Santiago, Chile}
\affiliation{Max-Planck-Institut f\"{u}r Astronomie, K\"{o}nigstuhl 17, 69117, Heidelberg, Germany}

\author[0000-0003-1837-3772]{Romane Le Gal}
\affiliation{Center for Astrophysics \textbar\ Harvard \& Smithsonian, 60 Garden St., Cambridge, MA 02138, USA}
\affiliation{IRAP, Universit\'{e} de Toulouse, CNRS, CNES, UT3, 31400 Toulouse, France}
\affiliation{Univ. Grenoble Alpes, CNRS, IPAG, F-38000 Grenoble, France}
\affiliation{IRAM, 300 rue de la piscine, F-38406 Saint-Martin d'H\`{e}res, France}

\author[0000-0002-7616-666X]{Yao Liu} \affiliation{Purple Mountain Observatory \& Key Laboratory for Radio Astronomy, Chinese Academy of Sciences, Nanjing 210023, China}

\author[0000-0002-7607-719X]{Feng Long}
\affiliation{Center for Astrophysics \textbar\ Harvard \& Smithsonian, 60 Garden St., Cambridge, MA 02138, USA}

\author[0000-0002-1637-7393]{Fran\c cois M\'enard}\affiliation{Univ. Grenoble Alpes, CNRS, IPAG, F-38000 Grenoble, France}

\author[0000-0002-7058-7682]{Hideko Nomura}
\affiliation{National Astronomical Observatory of Japan, 2-21-1 Osawa, Mitaka, Tokyo 181-8588, Japan}

\author[0000-0002-1199-9564]{Laura M. P\'erez} \affiliation{Departamento de Astronom\'ia, Universidad de Chile, Camino El Observatorio 1515, Las Condes, Santiago, Chile}

\author[0000-0001-8642-1786]{Chunhua Qi} \affiliation{Center for Astrophysics \textbar\ Harvard \& Smithsonian, 60 Garden St., Cambridge, MA 02138, USA}

\author[0000-0002-6429-9457]{Kamber R. Schwarz} \altaffiliation{NASA Hubble Fellowship Program Sagan Fellow}
\affiliation{Lunar and Planetary Laboratory, University of Arizona, 1629 E. University Blvd, Tucson, AZ 85721, USA}

\author[0000-0002-5991-8073]{Anibal Sierra} \affiliation{Departamento de Astronom\'ia, Universidad de Chile, Camino El Observatorio 1515, Las Condes, Santiago, Chile}

\author[0000-0003-1534-5186]{Richard Teague}\affiliation{Center for Astrophysics \textbar\ Harvard \& Smithsonian, 60 Garden St., Cambridge, MA 02138, USA}

\author[0000-0002-6034-2892]{Takashi Tsukagoshi} \affiliation{National Astronomical Observatory of Japan, 2-21-1 Osawa, Mitaka, Tokyo 181-8588, Japan}

\author[0000-0003-4099-6941]{Yoshihide Yamato}
\affiliation{Department of Astronomy, Graduate School of Science, The University of Tokyo, Tokyo 113-0033, Japan}

\author[0000-0002-2555-9869]{Merel L. R. van 't Hoff}
\affiliation{Department of Astronomy, University of Michigan, 323 West Hall, 1085 South University Avenue, Ann Arbor, MI 48109, USA}

\author[0000-0002-1566-389X]{Abygail R. Waggoner} 
\affiliation{Department of Chemistry, University of Virginia, Charlottesville, VA 22904, USA}

\author[0000-0003-1526-7587]{David J. Wilner}\affiliation{Center for Astrophysics \textbar\ Harvard \& Smithsonian, 60 Garden St., Cambridge, MA 02138, USA}

\author[0000-0002-0661-7517]{Ke Zhang}
\altaffiliation{NASA Hubble Fellow}
\affiliation{Department of Astronomy, University of Wisconsin-Madison, 475 N Charter St, Madison, WI 53706}
\affiliation{Department of Astronomy, University of Michigan, 323 West Hall, 1085 South University Avenue, Ann Arbor, MI 48109, USA}

\begin{abstract}

Planets form and obtain their compositions in dust and gas-rich disks around young stars, and the outcome of this process is intimately linked to the disk chemical properties. The distributions of molecules across disks regulate the elemental compositions of planets, including C/N/O/S ratios and metallicity (O/H and C/H), as well as access to water and prebiotically relevant organics. Emission from molecules also encodes information on disk ionization levels, temperature structures, kinematics, and gas surface densities, which are all key ingredients of disk evolution and planet formation models. The Molecules with ALMA at Planet-forming Scales (MAPS) ALMA Large Program was designed to expand our understanding of the chemistry of planet formation by exploring disk chemical structures down to 10 au scales. The MAPS program focuses on five disks -- around IM Lup, GM Aur, AS 209, HD 163296, and MWC 480 -- in which dust substructures are detected and planet formation appears to be ongoing. We observed these disks in 4 spectral setups, which together cover $\sim$50 lines from over 20 different species. This paper introduces the ApJS MAPS Special Issue by presenting an overview of the program motivation, disk sample, observational details, and calibration strategy. We also highlight key results, including discoveries of links between dust, gas, and chemical sub-structures, large reservoirs of nitriles and other organics in the inner disk regions, and elevated C/O ratios across most disks. We discuss how this collection of results is reshaping our view of the chemistry of planet formation.

\end{abstract}

\keywords{Astrochemistry}

\section{Introduction: the Chemistry of Planet Formation} \label{sec:intro}

Planets form in disks of dust and gas around young stars. The distributions of volatile elements and organics in these disks affect multiple aspects of planet formation \citep[see][for recent reviews]{Henning13,Dutrey14,Pontoppidan2014,Oberg21_Review}. Radial and vertical chemical gradients  may impact where in disks planets form, planet formation efficiencies,  planet elemental and organic compositions, and the isotopic ratios in planets and planetesimals. 
Furthermore, molecular line observations often provide the best, and sometimes our only, probes of disk characteristics relevant to planet formation such as surface density, ionization, temperature, metallicity, C/N/O/S ratios, and disk kinematics. The overall goal of the Molecules with ALMA at Planet-forming Scales (MAPS) Large Program is to use high-spatial resolution observations of a large number of molecular lines to promote a deeper understanding of the links between chemistry and planet formation in protoplanetary disks.

\begin{figure*}[htbp]
\begin{center}
\includegraphics[width=0.8\textwidth]{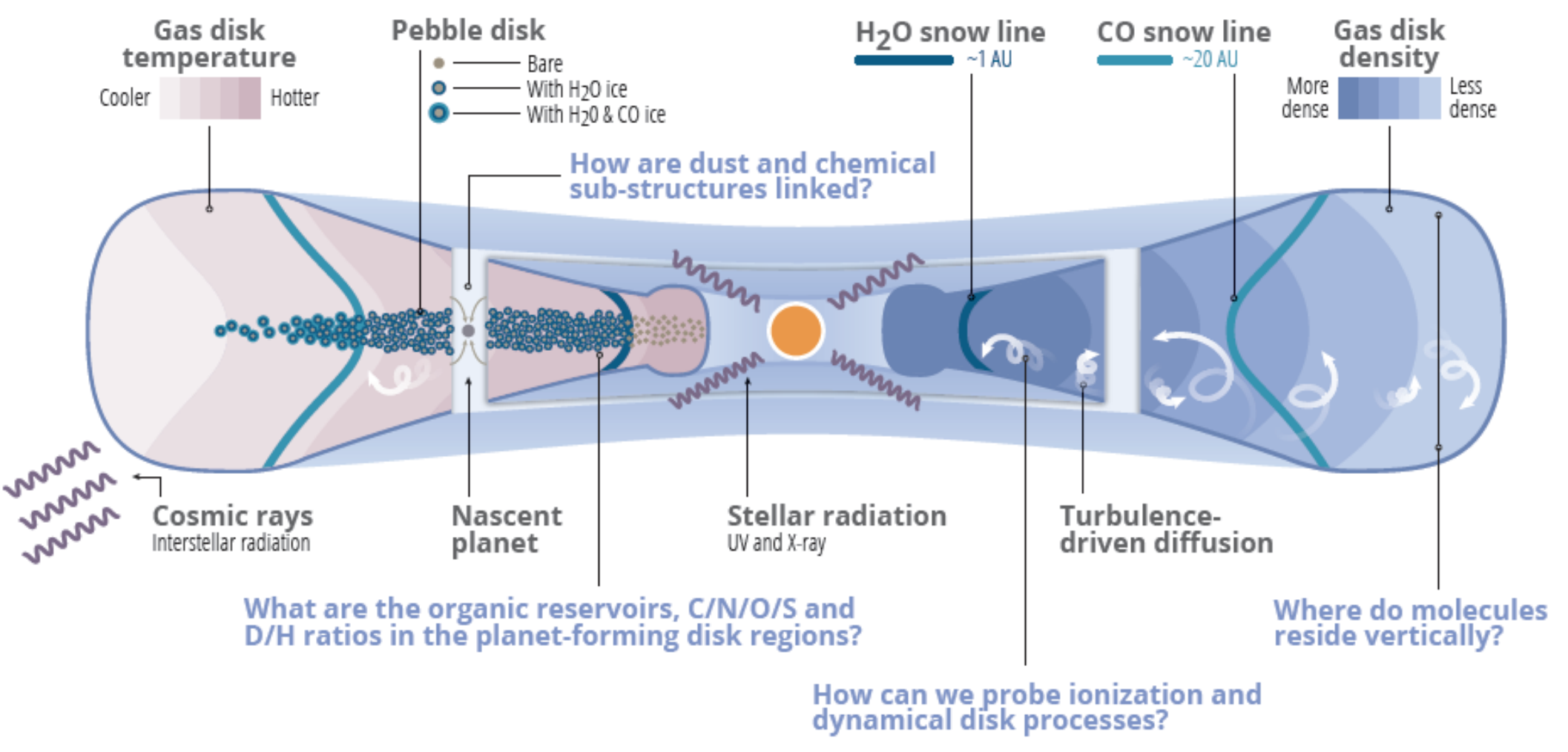}
\caption{Illustration of the protoplanetary disk structure, on a logarithmic scale, and its coupling to different gas and grain dynamical processes.  Note the temperature gradient inward and upward (pink shading), which results in a series of midplane snowlines (shown at typical locations for a disk around a T Tauri star) and 2D snow-surfaces reaching up into the disk atmosphere, as well as the inward and downward density gradient (blue shading). Disk surfaces are characterized by photon processes. Disk midplanes are by contrast cold and UV-poor, and the main volatile reservoirs (other than H$_2$ and He) are in icy grain and pebble mantles, especially exterior to the CO snowline. The four guiding questions used to design MAPS are written in blue. Image credit: K. Peek, adapted from \citet{Oberg21_Review}}
\label{fig:disk-cartoon}
\end{center}
\end{figure*}

The context within which MAPS was conceived is our current understanding of the structure, dynamics, and chemistry of protoplanetary disks, as illustrated in Fig.\ \ref{fig:disk-cartoon}.  Globally, protoplanetary disks are characterized by radial and vertical gradients in temperature, density, ionization, and radiation fields  \citep[e.g.,][]{vanZadelhoff01,Dartois03,Kamp04,Meijerink12,Woitke16}. These disk characteristics are important in their own right; ionization levels, surface densities, and temperature gradients are all key players in planet formation models \citep[e.g.,][]{Mordasini16,Baruteau16}. They also regulate the distribution, formation, and destruction of molecules in disks. For the purpose of introducing the disk chemistry most relevant to MAPS, we will treat these gradients as static. It is, however, important to note that protoplanetary disks are dynamical objects and that this can have a large impact on the distribution of molecules \citep[e.g.,][]{Willacy07,Semenov11,Akimkin13b,Rab17,Price20}. Globally, these accretion disks spread and deplete mass onto the star over time \citep{LyndenBell74,Hartmann16}. Within the disks, accretion flows transport gas and entrained grains both inward and outward, while grain settling and drift  transport solids towards the midplane and pressure maxima, and turbulence may mix gas and grains vertically and radially \citep{Weidenschilling93,Hartmann00,Birnstiel12}.

Of the disk properties illustrated in Fig. \ref{fig:disk-cartoon}, temperature gradients have long been of special interest. 
Radial temperature gradients are expected to produce a sequence of condensation fronts or snowlines, where abundant volatiles transition from gas to ice, in the planet-forming midplane \citep{Lewis74,Hayashi81,Qi13c,vantHoff17,Qi19}. These snowlines may impact the planet formation efficiency \citep[e.g.,][]{Lewis74}. At snowlines, dust coagulation properties change with changing grain compositions \citep{Dominik97,Guttler10,Wada13,Gundlach15,Pinilla17}, while rapid disintegration of pebbles crossing the snowline  result in traffic jams \citep{Birnstiel10}, and diffusive flows across snowlines result in large reservoirs of solids directly outside of the snowline location \citep{Stevenson88,Ciesla06,Ros13}. 
Snowlines may also regulate the elemental compositions of planets and planetesimals  \citep[e.g.,][]{Whipple72,Lewis74,Oberg11e,Piso16}. Because major carriers of common volatiles freeze out at different temperatures, C, O, N, and S are expected to deplete out of the gas at different disk radii, resulting in a sequential change in solid and gas composition. This should in turn translate into a dependence of planet core and atmosphere elemental composition on assembly location  \citep[e.g.,][]{Helling14,Cridland20}. It is worth noting that a combination of observations and theory suggest that the nature of the major carriers may change over time, especially due to the conversion of CO into other molecules  \citep{Favre13,Reboussin15, Yu17,Bosman18,Dodson-Robinson18,Schwarz18}. 

Condensation fronts also exist in the vertical dimension, resulting in two-dimensional snow-surfaces \citep{Aikawa02,vantHoff17,Qi19}, due to vertical temperature gradients. The vertical temperature structure is set by a combination of attenuating radiation from the central star, chemical feed-back on heating and cooling, and, in the inner disk, accretion \citep[e.g.,][]{dAlessio99,Woitke09}, but is also sensitive to dust growth and settling to the midplanes \citep[e.g.,][]{dAlessio06,Tilling12}.  The changing elemental ratios across the resulting snow-surfaces are expected to induce chemical gradients due to changing gas-phase and grain-surface elemental inventories. 

The vertical chemical structure is also strongly influenced by gradients in UV radiation and other ionizing agents \citep[e.g.,][]{Aikawa01,Bergin03,Semenov06,Nomura07,Willacy07,Dutrey07, Woitke09,Walsh10,Kamp10,Fogel11, Cazzoletti18}, because many chemical reactions are regulated by either UV photodissociation or ion-molecule reactions. The ionization structures of disks depend on a combination of  UV and X-ray disk surface fluxes and their radiative transfer, and on cosmic ray ionization \citep{Glassgold97,Cleeves14a,Woitke16,Drabek16,Rab18}. The total UV flux further depends on a combination of stellar UV radiation, UV from accretion shocks and external UV fields. Because of the important role of UV radiation, disk atmospheres have sometimes been described as analogs to photondominated regions (PDRs), where   different molecular photodissociation cross sections determine molecular emitting layers  \citep[e.g.,][]{vanDishoeck06,Cazzoletti18,Agundez18}.  

Millimeter and submillimeter astronomy has been used since the late 1980s to explore disk molecular lines and the underlying disk chemistry \citep[e.g.,][]{Weintraub89,Koerner93,Dutrey97, Kastner97,Thi04}. Spatially and spectrally resolved molecular line observations enable us to trace  radial disk chemical structures \citep[see ][for pioneering work]{Dutrey07,Qi08,Henning10}. Over the past cycles,  observations with Atacama Large Millimeter/submillimeter Array (ALMA) have been used to characterize the CO gas abundance, identify CO snowlines, and to map out gradients in other abundant volatiles and organic molecules \citep[e.g.,][]{Qi13c,Du15,Oberg15,Walsh16,Schwarz16, Bergin16,Miotello17, Zhang17,Cleeves18,Kastner18,Semenov18, Loomis18b,Zhang20a,LeGal19,vanTerwisga19,Tsukagoshi19,Rosotti21, Facchini21,Nomura21}.   Millimeter observations of edge-on disks  present the perhaps most straightforward path towards characterizing vertical chemical gradients \citep{Dutrey17,Louvet18,Teague20c,Podio20,vantHoff20}, but vertical chemical structures are also accessible in moderately inclined disks \citep{Pietu07,Semenov08,Rosenfeld13,Pinte18,Paneque21,Rich21},  such as the ones pursued in this program.

Most disk chemistry observations have  been confined to angular scales of $\gtrsim$0\farcs5 or 50--75~au (at a source distance of 100--150~pc), which has made it difficult to directly connect observed chemical structures with the material feeding planet formation. We note that the precise molecular inventory on small scales is especially interesting when considering how disk chemistry may impact the volatile composition of temperate, Earth-like planets where the subsequent chemistry may result in the development of life. While we do not know how life originated on Earth, nitriles and other small, reactive organics are implicated in several origins of life scenarios \citep{Powner09,Patel15,Pearce17} and their distributions in the inner regions of disks are of special interest. 
To asses the molecular inventory in the disk regions most commonly associated with planet formation \citep{Pollack96} requires high resolution observations.

Higher spatial resolution observations are also needed to explore  the relationship between dust and chemical substructures. One of the great discoveries in protoplanetary disk science in the past decade is that disks frequently display substructures -- rings, gaps, spirals, and clumps -- in millimeter continuum emission, tracing the distribution of small pebbles \citep[e.g.,][]{Brogan15,Andrews16,Andrews18,Huang18b, Long18,Andrews20, Cieza21}. The observed gaps have often been connected to ongoing planet formation \citep[e.g.,][]{Flock15,Fedele17,Keppler19}, though other plausible explanations exist as well \citep[see][for a review]{Andrews20}. Existing observations suggest that the distribution of dust can be important for explaining and predicting the distribution of different molecules.  In particular, the edges of pebble disks are sometimes associated with dramatic changes in the molecular line emission pattern \citep[e.g.,][]{Oberg15b,Bergin16}. Similarly, observations of bright HCO$^+$ emission in the large gaps associated with transition disks indicate high gas-to-mm-grain ratios compared to their surroundings, which should result in chemically distinct disk regions \citep[e.g.,][]{Drabek16,Huang20}. Together with model predictions \citep[e.g.,][]{vanderMarel18,Facchini18,Alarcon20a}, these observations suggest that dust and pebble gaps may cause observable chemical changes. This chemical response to the opening of pebble gaps in the disk will determine the local chemical environment within which massive gap-opening planets form.
 
In addition to providing direct measures of chemical abundances, molecular lines provide constraints on the elemental composition, surface density, temperature, ionization, and kinematics of disk gas \citep[some recent examples can be found in][but see also the reviews above for references to earlier work]
{vanderMarel16, Dutrey17,Miotello17,Flaherty17, Huang18,Pinte18,Calahan20}. 
Some of these molecular probes work because some molecular abundances sensitively depend on C/N/O/S ratios, 'metallicity' (C/H and O/H ratios), disk ionization, UV radiation, density, temperature, and D/H levels. Elemental ratios and gas metallicity are, for example, expected to affect the ratios of common disk molecules such as C$_2$H/CO, CS/SO, and HCN/CO as well as the abundances of more complex molecules like CH$_3$CN \citep{Du15,Cleeves18, Semenov18,Miotello19,LeGal19b}, while other molecular pairs are proposed to be sensitive to ionization (HCO$^+$/CO), or UV (CN/HCN) \citep{Bergin03,Teague15}. Molecular isotopologues are proposed to be especially powerful diagnostics \citep{Ceccarelli14}. Deuterium enhancements are commonly used to trace the origins of Solar System volatiles, but these conclusions currently depend largely on theoretical expectations of D/H levels in volatiles in disks \citep[e.g.,][]{Aikawa99,Aikawa01,Willacy07,Willacy09,Cleeves14,Cleeves16b}. A few resolved observational studies have helped elucidate important drivers for deuterium chemistry  \citep[e.g.,][]{Oberg12,Mathews13, Huang17,Salinas17, Carney18,Oberg21_TWHya}, but the distribution of D/H in the planet-forming disk regions is still largely unknown.

Molecular emission lines are also potential probes of disk gas masses and gas surface densities \citep[e.g.,][]{Williams14,Miotello16, Molyarova17}, and can be used to constrain what types of planetary system could form in disks \citep[e.g.,][]{Rab20}. 
Molecular emission may also constitute our best tools to characterize how and where planets form in disks. Observations of molecular lines are key to characterizing the distribution and dynamics of gas in disks \citep[e.g.,][]{Dartois03,Isella07,Rosenfeld12a}. More specifically, spatially and spectrally resolved lines can be used to identify chemical,  and kinematic signatures of planets in the making \citep[e.g.,][]{Kanegawa15,Perez15, Cleeves15b,Pinte18,Teague18a,Dong19,Tsukagoshi19,Nomura21}. This is an exciting new development, since direct planet detections in disks remain rare \citep{Keppler18,Wang21}.  

With this background in mind, we designed the ALMA Large Program Molecules with ALMA on Planet-forming Scales (MAPS) to address the following goals (see also Fig.\ \ref{fig:disk-cartoon}):

\begin{enumerate}
    \item to assess the relationship between dust substructures and gas and chemical substructures in disks,
    \item to constrain the emitting heights of observed molecules and assess how well we probe planet-forming layers,
    \item to take astrochemistry studies into the planet forming disk regions and address the C/N/O/S ratios, deuterium fractionation, and organic compositions at scales down to 10~au, and
    \item to constrain disk dynamics, temperature, gas surface densities, and ionization across disks, especially in disk gaps and rings, using a variety of molecular probes.
\end{enumerate}

\vspace{-2mm}

This paper introduces the Special Issue of ApJS reporting our results from MAPS, including both analysis of the ALMA data  and models aimed at providing an interpretive framework. In particular, the purpose of this paper is to present the MAPS observational program, to  introduce the other papers in the special issue, and to discuss how the results of the individual MAPS papers are advancing our understanding of disk chemistry and its relationship to planet formation. In \S\ref{sec:sample} we describe and motivate the disk sample and the lines targeted with MAPS. \S\ref{sec:obs} presents the observations, our calibration strategy, and brief summary of the imaging strategy. The latter is described in detail in \citet{Czekala21}.  \S\ref{sec:res} provides an overview of the results presented in the MAPS papers in this special issue of ApJS, while \S\ref{sec:disc} discusses how this ensemble of results has changed our view of the chemistry of planet formation, and \S\ref{sec:conc} provides a brief summary and some concluding remarks.

\section{MAPS Disks and Lines} \label{sec:sample}

\subsection{MAPS Disk Sample}

\begin{figure*}[ht!]
\centering
\includegraphics[scale=0.7]{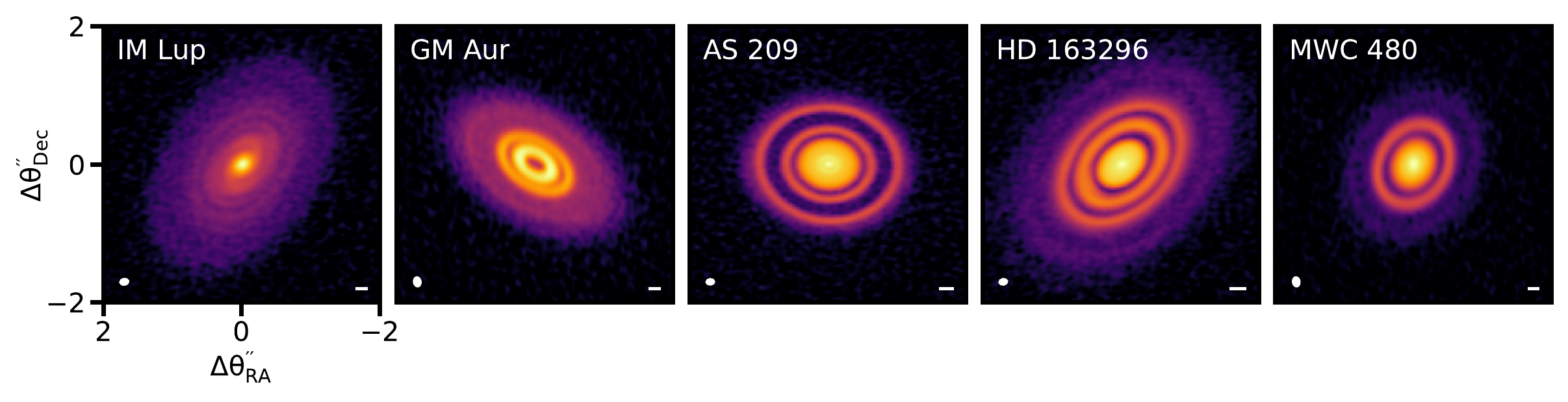}
\caption{220~GHz continuum images from MAPS  displaying the diversity of continuum structure present at the same spatial resolutions as probed by the MAPS molecular lines. The images are shown with an {\it arcsinh} color stretch, which accentuates low surface brightness features. The synthesized beams are indicated in the bottom-left corner and a 20 au scale bar in the bottom-right corner of each panel. See \citet{Law21_radprof} and \citet{Sierra21} for a detailed presentation and analysis of these images. \label{fig:sources}}
\end{figure*}

The MAPS sample consists of five disks which were selected to achieve a diversity of dust substructures and stellar properties, and to ensure line detections. 
In brief, we selected the sample based on the following considerations:

\begin{enumerate}
    \item The sample should include both T Tauri and Herbig Ae stars to explore the role of the stellar luminosity and radiation field on disk chemistry.
    \item The sample should include different kinds of dust substructures to enable identifications of links between dust and chemical substructures.
    \item No disks should be substantially obscured by cloud material or include substantial envelope emission that would complicate the interpretation of line observations.
    \item All disks should have been previously observed in at least a subset of the targeted molecular lines.
\end{enumerate}

Practically, we selected the MAPS disk sample by first identifying sources that had resolved continuum observations at $\sim$0\farcs1 or better with ALMA, and that had been included in disk chemistry related observational programs with the Plateau de Bure Interferometer (PdBI), the Northern Extended Millimeter Array (NOEMA), the Submillimeter Array (SMA), or ALMA \citep{Dutrey07,Oberg10c,Oberg11a,Huang17,Bergner18}. We first explored the disks observed within the ALMA Large Program DSHARP \citep{Andrews18}, and five of the DSHARP disks met our criteria. Of these, we selected two disks around T Tauri stars (IM Lup and AS 209) and one around a Herbig Ae star (HD 162396) based on their diversity of dust substructures. We supplemented this sample with an additional disk around the Herbig Ae star MWC 480, whose continuum was characterized by \citet{Long18}, and with a transition disk (a disk with a large inner gap or cavity) around the T Tauri star GM Aur, which was recently observed at high resolution by \citet{Huang20}.

The resulting MAPS sample consists of three T Tauri (IM Lup, AS 209, and GM Aur) and two Herbig Ae (HD 163296 and MWC 480) star-disk systems. All disks have dust gaps and rings, but the nature of these gaps and rings vary substantially between the five disks \citep{Andrews18,Huang18b,Long18,Huang20}, as illustrated in Fig.\ \ref{fig:sources}. Additional dust features include a large central gap in the GM Aur disk, spiral arms in the IM Lup disk \citep{Huang18c}, a clear dust asymmetry in the HD 163296 disk \citep{Isella18}, and faint extended dust emission beyond the bright outermost disk ring towards GM Aur, IM Lup, and HD~163296. The sample star and disk properties are summarized in Table \ref{tab:sources}.  In particular, Table \ref{tab:sources} lists stellar spectral types, masses, effective temperatures, luminosities, approximate ages, accretion rates, and Gaia distances; disk inclinations and position angles; and source velocities based on millimeter spectral lines. We note that ages of young stars are notoriously difficult to constrain, and often vary by a factor of 2 or more between different studies. The values noted in Table \ref{tab:sources} constitute recent estimates for the MAPS sources, but may be further revised by future studies. Mass accretion rates also vary  between different literature sources, which at least in part is due to substantial accretion variability on short time scales \citep{Mendigutia13,Ellerbroek14,Ingleby15}. This uncertainty in average accretion rates makes it difficult to use literature accretion rate values to explain disk chemical properties, since bursts in accretion may have substantial impact on the chemistry \citep{Rab17}.  
Pertinent information about individual sources is described below.

\begin{deluxetable*}{lccccccccccc}
\tablecaption{Stellar and Disk Properties \label{tab:sources}}
\tablehead{
\colhead{Source} & \colhead{Spectral Type} &
\colhead{dist.\tablenotemark{a}} & \colhead{incl} & \colhead{PA} & \colhead{T$_{\rm eff}$} &\colhead{$L_{\rm *}$\tablenotemark{b}}  &\colhead{Age\tablenotemark{c}}&\colhead{$M_*$\tablenotemark{d}}  &\colhead{log$_{10}$($\dot{M}$)\tablenotemark{b}} &\colhead{$v_{\rm sys}$}  &\colhead{References}\\
\colhead{} & \colhead{} 
&\colhead{[pc]}&\colhead{[$^{\circ}$]}&\colhead{[$^{\circ}$]}&\colhead{[K]} &\colhead{[$L_{\rm \odot}$]} & \colhead{[Myr]} &\colhead{[$M_{\rm \odot}$]} &\colhead{[$M_{\rm \odot}$ yr$^{-1}$]} & \colhead{[km~s$^{-1}$]}
}
\startdata
IM Lup  &K5 
&158    &47.5   &144.5  &4266   &2.57   & $\sim1$ &1.1   &$-$7.9   &4.5    &1,2,3,4,5,12\\
GM Aur  &K6 
&159 &53.2  &57.2  &4350   &1.2     &$\sim3$--10& 1.1   &$-$8.1   &5.6  &1,6,7,8,9,10,11\\
AS 209  &K5 
&121    &35.0  &85.8  &4266   &1.41   &$\sim1-2$& 1.2    &$-$7.3   &4.6    &1,2,12,13,14\\
HD 162396   &A1   
& 101 &46.7   &133.3  &9332   &17.0 &$\gtrsim6$ &2.0  &$-$7.4   &5.8  &1,2,12,15,16\\
MWC 480 &A5 
&162   &37.0 &148.0    &8250   &21.9  &$\sim7$ &2.1   &$-$6.9   &5.1    &1,17,18,19,20,21\\
\enddata
\tablenotetext{a}{Gaia DR2 distances, which were adopted since DR3 distances were not available at the start of the project. DR3 distances were checked for all sources and have differences of at most a few au for IM~Lup, GM~Aur, and MWC~480, which does not alter our results.}
\tablenotetext{b}{Stellar luminosities and accretion rates extracted prior to Gaia have been updated with the new distances following \citet{Andrews18} for IM~Lup, AS~209, and HD~163296 and \citet{Macias18} for GM~Aur. Due to their large relative uncertainties, those for MWC~480 have not been rescaled. The accretion rate of GM~Aur is variable \citep{Ingleby15}, so we adopted an average value, following \citet{Macias18}.}
\tablenotetext{c}{The stellar ages are uncertain by at least a factor of two and should be considered provisional.}
\tablenotetext{d}{All stellar masses have been dynamically determined as described in \citet{Teague21}.}
\tablecomments{References are 1. \citet{Gaia18}; 2. \citet{Huang18b}; 3. \citet{Alcala17}; 4. \citet{Pinte18b}; 5. \citet{Mawet12}; 6. \citet{Huang20}; 7. \citet{Macias18}; 8. \citet{Espaillat10}; 9. \citet{Kraus09}; 10. \citet{Beck19}; 11. \citet{Ingleby15}; 12. \citet{Andrews18}; 13. \citet{Salyk13}; 14. \citet{Huang17}; 15. \citet{Fairlamb15}; 16. \citet{Teague19};  
17. \citet{Liu19}; 18. \citet{Montesinos09}; 19. \citet{Simon19}; 20. \citet{Pietu07}; 21. \citet{Mendigutia13}}
\end{deluxetable*}

{\it IM Lup.} IM Lup is located in the Lupus star forming region and likely hosts the youngest ($\sim$1 Myr) disk in the sample \citep{Mawet12}, though a range of age estimates exist in the literature. It is approximately Solar-mass and can be viewed as a young analog to the Sun. Its disk was observed by \citet{vanKempen08} and characterized by \citet{Pinte08} and \citet{Panic09}. Despite its youth, the IM Lup disk shows clear evidence for grain growth \citep{Lommen07}.   It stands out among other Lupus sources due to its disk mass, estimated to be 0.1-0.2 M$_{\rm \odot}$ based on continuum (spectral energy distribution, scattered light images, and resolved millimeter images), and CO multi-isotopologue, multi-line data \citep{Pinte08,Cleeves16c}. Such a high disk mass would typically be associated with a high accretion rate \citep{dAlessio99}, and while early studies failed to find signs of accretion \citep{Padgett06,Gunther07}, more recent studies have shown that IM Lup is a rather typical T Tauri accretor \citep{Alcala17}. 

The IM Lup disk is not only massive, but also spatially very extended in both dust and gas. The pebble disk has a radius of $\sim$264~au, which was the largest one observed within the DSHARP project \citep{Huang18b}. It also stands out for its size in other surveys where it was included, and when compared to derived disk size scaling laws \citep{Andrews18,Hendler20}. The CO disk extends far beyond the dust pebble disk \citep{Panic09}, out to $\sim$900~au, though the outer CO disk regions appear fluffy or envelope-like \citep{Panic09,Cleeves16c}. IM Lup has also been observed in scattered light \citep{Pinte08}, revealing a beautifully flared, multi-ringed disk in small dust grains \citep{Avenhaus18}. IM Lup's disk spiral in millimeter emission is noteworthy, since in the DSHARP sample spirals around single stars were otherwise rare \citep{Huang18c}.

The molecular inventory of IM Lup beyond CO was first catalogued by \citet{Oberg11a} using the SMA as a part of the DISCS (Disk Imaging Survey of Chemistry with SMA). With ALMA, IM Lup has been observed in CO isotopologues ($J=2-1$ and $3-2$) \citep{Cleeves16c, Pinte18b}, H$^{13}$CO$^+$ and DCO$^+$ 3--2 \citep{Oberg15b,Cleeves16c}, DCN and HCN 3--2 \citep{Huang17}, H$_2$CO $4_{04}-3_{03}$ \citep{Pegues20}, C$_2$H 3--2 \citep{Bergner19}, and N$_2$H$^+$ 3--2 \citep{Seifert21}. Notably, IM Lup shows a stunning double-ring structure in DCO$^+$ 3-2, which has been attributed to a change in the gas-phase C and O abundances at the pebble disk edge \citep{Oberg15b}. It is undetected in complex nitriles \citep{Bergner18}. The IM Lup disk was modeled in detail in \citet{Cleeves16c} and \citet{Cleeves18}, and found to be underabundant in CO by about a factor of 20 compared to molecular clouds, indicative that both oxygen and carbon are missing from the gas-phase. By contrast, nitrogen does not appear to be depleted, resulting in elevated gas-phase N/O. 

 IM Lup was searched for water vapor in the far-IR with {\it Herschel}  \citep{vanDishoeck11,Du15}. While there was no detection, indicative of water and O depletion, the upper limits were less constraining compared to some other disks, and this data has not yet been used to quantify the IM Lup H$_2$O reservoir. IM Lup was included in {\it Spitzer} surveys of disk volatiles, but was initially undetected in most molecules (H$_2$O, OH, HCN, and C$_2$H$_2$) except for CO$_2$, indicating that its inner disk atmosphere is chemically poor \citep{Pontoppidan10,Salyk11}. However, IM Lup was revisited by \citet{Banzatti17}, who detected weak OH and H$_2$O emission, but in the case of H$_2$O only at longer (33 $\mu$m) IR wavelengths. They suggested that IM Lup was at an intermediate stage of water depletion and still retains some water vapor in its inner disk.

{\it GM Aur.} GM Aur is a $\sim$3--10~Myr T Tauri star in the Taurus-Auriga star-forming region  with a so called `transition' disk, i.e., a disk with a large inner gap or cavity. \citet{Strom89} noted it for its weak near-IR excess compared to other T Tauri stars, and speculated that this and similar sources were disks in transition, which lead to the name `transition disks'. GM Aur's large dust cavity of $\sim$35~au in radius was first inferred from SED modeling \citep{Marsh92,Calvet05,Espaillat10} and later directly imaged using millimeter observations \citep{Hughes08}. More recently \citet{Macias18} and \citet{Huang20} used high-resolution ALMA observations to reveal that the outer disk consists of nested rings. \citet{Huang20} also identified a small disk inside the gap, and observed that the outer rings are surrounded by low-intensity continuum emission extending to 270~au. The mass of the GM Aur disk was recently constrained using {\it Herschel} observations of HD to 0.03-0.2 M$_\odot$ \citep{McClure16}.

GM Aur has also been observed in multiple gas tracers at millimeter wavelengths. Both \citet{Dutrey08} and \citet{Hughes09} found that the CO emission was not entirely Keplerian towards the GM Aur disk, and hypothesized that the disk is warped by a planet. A range of other common molecular tracers were clearly detected  with the SMA \citep{Oberg10c}, revealing that GM Aur is bright in molecular lines associated with photochemistry and small organics. In contrast with all other disks in this program, GM Aur was not part of any early ALMA disk chemistry surveys and there are therefore fewer sub-arcsecond-resolution molecular observations. Notable exceptions are H$_2$CO, DCO$^+$, and N$_2$H$^+$ \citep{Qi19, Pegues20}. N$_2$H$^+$ was further used to constrain the CO and N$_2$ snowline locations  to 48 and 78~au \citep{Qi19}. In addition, \citet{Huang20} observed the GM~Aur disk in HCO$^+$ $J=3-2$ at a higher spatial resolution, $\sim$0\farcs1, and found that the radial emission profile has substantial substructure, most of which could be attributed to continuum absorption.

GM Aur was included in {\it Herschel} spectral line surveys and found to be somewhat depleted in oxygen ([O~I]) \citep{Keane14}, not detected in water \citep{Du17}, and not detected in [C~II] \citep{Howard13}. GM Aur was not included in the large {\it Spitzer} surveys of disk molecular lines due to its low IR flux \citep{Pontoppidan10}, and its inner disk molecular content is therefore unknown.

{\it AS 209.} AS 209 is a young ($\sim$1--2 Myrs) T Tauri star in the Ophiuchus star-forming region. The global disk structure was modeled by \citet{Qi19} and found to have a thin isothermal midplane layer and an inner wall. The disk is highly structured in millimeter continuum emission with at least 7 nested rings \citep{Andrews09,Fedele18,Andrews18,Huang18,Guzman18}, some of which have been associated with ongoing planet formation \citep{Guzman18, Favre19}. In addition, observations at mm and cm wavelengths shows clear evidence for both grain growth and pebble-size dependent grain drift \citep{Perez12,Tazzari16}. Observations in scattered light also indicate the presence of three rings \citep{Avenhaus18}. 

Based on CO isotopologue observations, the radial gas surface density and/or CO abundance profiles are highly structured as well. \citet{Huang16} first showed that C$^{18}$O has a substantial gap and a ring close to the millimeter dust edge. This observation was confirmed by \citet{Favre19} using higher resolution and higher sensitivity observations, which enabled them to associate the C$^{18}$O gap with a dust gap. Observing the main CO isotopologue, \citet{Guzman18} found multiple depressions in the CO radial profile that could be related to both peaks or gaps in the dust emission radial profile.  Furthermore, \citet{Teague18b} found that the continuum and NIR structures are associated with gas surface density perturbations, derived from the rotational velocity of $^{12}$CO.

AS 209 is known to have a rich molecular line inventory at millimeter wavelengths \citep{Oberg11a, Huang17,Bergner18,Bergner19}. Most of the millimeter molecular emission  appears to be depleted towards the disk center and whether this is an effect of dust opacity, excitation, C/N/O abundance, or gas surface density has not been resolved. This is addressed within MAPS in \citet{Bosman21_XV}.  The inner disk chemistry of AS 209 has not been well characterized, and there are only upper limits from {\it Herschel} on the water reservoir \citep{Du17}.

{\it HD 163296.} HD 163296 is  one of the most well-studied Herbig Ae star-disk system at millimeter wavelengths due to its relative proximity and massive disk; it is estimated to contain 0.1--0.5 M$_{\rm \odot}$ \citep{Isella07,Tilling12,Muro-Arena18,Powell19,Kama20}. The age of HD~163296 is somewhat unclear, but most studies suggest it is at least 6~Myrs old \citep{Fairlamb15, Wichittanakom20}. The disk presents multiple  features that suggest the presence of planets and ongoing planet formation, including dust rings, azimuthal asymmetries, deviations from Keplerian velocities due to gas pressure variations, `kinks' in the CO emission, and meridional flows \citep{Isella16,Isella18,Pinte18,Teague18a,Teague19,Pinte20,Rodenkirch21}. The three large circular gaps in millimeter continuum at 45, 87, and 140~au are especially visible in Fig.\ \ref{fig:sources}. The HD~163296 disk has been  observed in scattered light at optical and infrared wavelengths, revealing substantial radial substructure also at these shorter wavelengths \citep{Monnier17,Muro-Arena18,Rich20}.

HD 163296 has been observed in CO and other molecular tracers  at millimeter wavelengths as part of a number of single-dish, SMA, and ALMA projects \citep[e.g.,][]{Thi04,Qi11,Oberg11a,Rosenfeld13,Klaassen13,Mathews13,deGregorio13, Flaherty15,Huang17,Salinas17,Guzman18,Booth19,Bergner19,Notsu19, Pegues20,Zhang20a}.  Its popularity in previous surveys can be explained by its bright molecular lines, not only in CO isotopologues, but also in HCN, HCO$^+$, deuterated species, and larger organics.  HD~163296 was subject to a deep search for CH$_3$OH, which yielded an abundance upper limit well below what has been seen towards the nearby TW Hya disk, indicative of a hostile environment for some families of organic molecules \citep{Carney19}. The CO snowline was first estimated by \citet{Qi11}, and then revised by \citet{Qi15} to 75~au, assuming a 101~pc (Gaia) distance.      

 Similar to most disks around Herbig Ae stars, the HD 163296 disk is relatively line poor at  IR wavelengths \citep{Pontoppidan10,Salyk11}, which may simply be a result of high continuum flux levels \citep{Antonellini16}. H$_2$O and OH have, however, been detected in the HD 163296 disk both at mid-IR (33 $\mu$m) and far-IR wavelengths  \citep{Fedele12,Banzatti17}, revealing a disk atmosphere that is not completely dry. HD 162396 has also been observed, though not detected, in high-J CO lines with {\it Herschel} \citep{Du17}, which together with continuum data, and [O~I] and low-J CO line detections have provided constraints on the disk temperature structure and atmospheric gas to dust ratio \citep{Tilling12}.   The inner disk also appears  asymmetric, potentially hosting a large-scale vortex, which may impact its chemistry \citep{Varga21}.

{\it MWC 480.} MWC 480 is a $\sim$7 Myr old Herbig Ae star in the Taurus-Auriga star-forming region \citep{Montesinos09}. Its disk has been characterized through millimeter continuum observations \citep{Pietu06}, and recent 0\farcs1 observations with ALMA show that the disk has an outer dust gap and ring, as well as low-level millimeter emission extending out to $\sim$200~au \citep{Long18,Liu19}. Scattered light images of the disk reveal that the smaller grains are similarly extended, but do not present any substructure \citep{Kusakabe12}.

The gas structure of the MWC 480 disk has been probed using CO observations, and the outer disk exhibits a vertical temperature gradient \citep{Pietu07}. \citet{Hughes08} showed that the CO gas is much more extended than millimeter pebbles. 

MWC 480 was included in both the Chemistry in Disks (CID) and DISCS programs and has therefore had its disk chemistry extensively explored at medium spatial resolution at both 1 and 3~mm \citep{Dutrey07,Oberg10c,Henning10,Dutrey11}. These early observations showed that while the disk is very bright in some lines, e.g.,\ HCO$^+$ and HCN, it is weak in others, including N$_2$H$^+$ and H$_2$CO \citep{Dutrey07,Pegues20}. MWC 480 was the first disk where CH$_3$CN was detected \citep{Oberg15Natur}, the most complex molecule seen in disks so far, and it has since been probed at $\sim$0\farcs5 resolution in simple and complex nitriles,  deuterated molecules, H$_2$CO, and sulfur-bearing species \citep{Guzman15,Huang17, Bergner18,LeGal19,Pegues20}. Most recently MWC 480 has been the target of an unbiased spectral line survey \citep{Loomis20}, which resulted in detections of several new disk molecules, including C$_2$D and H$_2$CS.

The MWC 480 water content was probed in depth by {\it Herschel} as a part of the Water In Star-forming regions with {\it Herschel} key program\citep{vanDishoeck11}, but was not detected \citep{Du17}. A stacked detection was obtained when combining the MWC 480 data with three other disks, but its implications for the MWC 480 water budget is difficult to ascertain \citep{Du17}. The inner disk chemistry of MWC~480 is unknown.

\subsection{Line targets}

\begin{deluxetable*}{llcccccccccc}
\tablecaption{Lines targeted at 3~mm (ALMA Band 3)}
\label{tab:lines-b3}
\tabletypesize{\scriptsize}
\tablehead{
\colhead{Molecule}  &\colhead{QN}   & \colhead{Rest freq.}    & \colhead{Log$_{10}(A_{\rm ij} / \rm{s}^{-1})$} & \colhead{$g_{\rm u}$} &
\colhead{$E_{\rm u}$}   &\colhead{Cat.$^1$} &\multicolumn{5}{c}{Detected}\\
  \colhead{} &  \colhead{} &\colhead{[GHz]}   & \colhead{}  & \colhead{} & 
\colhead{[K]}   & \colhead{}    &\colhead{IM Lup}   &\colhead{GM Aur}   &\colhead{AS 209}    &\colhead{HD 163296}    &\colhead{MWC 480}
}
\tablecolumns{12}
\startdata
{\bf B3-1}   \\
HC$^{15}$N &J=1--0   &86.054966 &$-$4.6569&3   &4.1   &CDMS   &N&N&N&N&N\\
H$^{13}$CN$^\dagger$& J=1--0   & 86.339921 &$-$4.6526&9&4.1&CDMS   &N&(Y)$^*$&N&N&N\\ 
H$^{13}$CO$^+$ &J=1--0   &86.754288 & $-$4.4142 & 3 & 4.2 & LAMDA  &(Y)$^*$  &Y  & (Y)$^*$ & Y & (Y)$^*$\\
C$_2$H  &N=1--0, J=$\frac{3}{2}$--$\frac{1}{2}$, F=2--1    &87.316898   &$-$5.6560 & 5& 4.2 & CDMS  &Y&Y&Y&Y&Y\\
&  N=1--0, J=$\frac{3}{2}$--$\frac{1}{2}$, F=1--0 &87.328585   &$-$5.7367 &3 &4.2  &CDMS   &Y&Y&Y&Y&Y\\
HCN$^\dagger$    &J=1--0, F=1--1  &88.630416  & $-$4.6184 & 3 & 4.3   & CDMS  &Y&Y&Y&Y&Y\\
&J=1--0, F=2--1 &88.631848 & $-$4.6185 & 5 & 4.3 & CDMS&Y&Y&Y&Y&Y\\
&J=1--0, F=0--1    &88.633936 & $-$4.6184 & 1 & 4.3 & CDMS&Y&Y&Y&Y&Y\\
HCO$^+$    &J=1--0   &89.188525  & $-$4.3715 & 3 & 4.3 & CDMS  &Y&Y&Y&Y&Y\\
HC$_3$N    &J=11--10 &100.076392 & $-$4.1096 & 23 & 28.8 & CDMS   &N&Y&Y&Y&Y\\
H$_2$CO    &(J$_{\rm{K}_{\rm{a}}, {\rm{K}_{\rm{c}}}}$)=$6_{15}$--$6_{16}$ &101.332991   &$-$5.8037&39&87.6&CDMS&N&N&N&N&N\\
\hline
{\bf B3-2}\\
CS  &J=2--1    &97.980953   & $-$4.7763  &5& 7.1  & CDMS   &Y&Y&Y&Y&Y\\
C$^{18}$O  &J=1--0   &109.782173  & $-$7.2030 & 3 & 5.3 & LAMDA   &Y&Y&Y&Y&Y\\	 
$^{13}$CO  &J=1--0   &110.201354   & $-$7.2011 & 3 & 5.3 & LAMDA  &Y&Y&Y&Y&Y\\
CH$_3$CN    &J=6--5, K=5  &110.330345 & $-$4.4697  & 26 & 197.1 & CDMS&N&N&N&N&N\\
&J=6--5, K=4 &110.349471 & $-$4.2098 & 26 & 132.8 & CDMS &N&N&N&N&N\\
&J=6--5, K=3 &110.364354  & $-$4.0792 & 52 & 82.8  & CDMS &N&N&N&N&N\\
&J=6--5, K=2 &110.374989 & $-$4.0054 & 26 & 47.1  & CDMS &N&(Y)$^*$&Y&N&N\\
&J=6--5, K=1 &110.381372 & $-$3.9664 & 26 & 25.7  & CDMS &N&(Y)$^*$&Y&(Y)$^*$&Y\\
&J=6--5, K=0 &110.383500 & $-$3.9542 & 26 & 18.5  & CDMS &N&(Y)$^*$&Y&(Y)$^*$&Y\\  
C$^{17}$O  &J=1--0, F=$\frac{3}{2}$--$\frac{5}{2}$  &112.358777  &$-$7.1739 & 4 & 5.4 & CDMS  &  (Y)$^*$& Y & (Y)$^*$& Y & Y\\ 
&J=1--0, F=$\frac{7}{2}$--$\frac{5}{2}$ &112.358982&$-$7.1739 & 8 & 5.4 &  CDMS  &  (Y)$^*$& Y &  (Y)$^*$& Y & Y \\
&J=1--0, F=$\frac{5}{2}$--$\frac{5}{2}$ &112.360007 &$-$7.1739 & 6 & 5.4 &  CDMS  &  (Y)$^*$& Y &  (Y)$^*$& Y & Y\\
CN& N=1--0, J=$\frac{3}{2}$--$\frac{1}{2}$ F=$\frac{3}{2}$--$\frac{1}{2}$   &113.488120   &$-$5.1716& 4 & 5.4 & CDMS  &Y&Y&Y&Y&Y\\
&N=1--0, J=$\frac{3}{2}$--$\frac{1}{2}$ F=$\frac{5}{2}$--$\frac{3}{2}$&   113.490970     &$-$4.9236& 6 & 5.4 & CDMS  &Y &Y & Y & Y & Y \\
&N=1--0, J=$\frac{3}{2}$--$\frac{1}{2}$ F=$\frac{1}{2}$--$\frac{1}{2}$   &113.499644     &$-$4.9735& 2 & 5.4 & CDMS  & Y& Y& Y& Y& Y \\
&N=1--0, J=$\frac{3}{2}$--$\frac{1}{2}$ F=$\frac{3}{2}$--$\frac{3}{2}$  &113.508907      &$-$5.2848& 4 & 5.4 & CDMS  & Y& Y& Y& Y& Y\\
\enddata
\tablenotetext{}{$^1$ Spectroscopic data are compiled from the CDMS \citep{Muller01, Muller05,Endres16}; JPL \citep{Pickett98}; and LAMBDA \citep{Schoier05} catalogues. See individual MAPS papers for complete spectroscopic references for each line.}
\tablenotetext{}{* Tentatively detected with a 3--5$\sigma$ significance \citep{Aikawa21,Cataldi21,Ilee21,Zhang21}.}
\tablenotetext{}{$\dagger$ See \citet{Cataldi21} for the full list of hyperfine components.}
\end{deluxetable*}

Within MAPS, we observed the five disks in four spectral set ups (Tables \ref{tab:lines-b3} and \ref{tab:lines-b6}). These were selected to address the largest number of science goals in the smallest number of spectral settings. We mainly focused on molecules that had been previously detected in disks, and all lines targeted in ALMA Band 6 had been detected in at least some of the disks in the sample. By contrast, the majority of lines targeted in ALMA Band 3 had not been observed in any disks prior to MAPS. 

The full set of molecules includes tracers of gas structure, mass and kinematics (CO isotopologues), C/N/O/S ratios (C$_2$H, HCN, CO isotopologues, and CS), disk organic inventories and chemistry (C$_2$H, HCN, H$_2$CO, $c$-C$_3$H$_2$, HC$_3$N, CH$_3$CN), deuterium fractionation (DCN and N$_2$D$^+$), disk ionization (HCO$^+$, H$^{13}$CO$^+$, and N$_2$D$^+$), and photochemistry (CN and C$_2$H). A substantial subset of species are covered in at least two  transitions with different upper energy levels -- $^{13}$CO, C$^{18}$O, HCN, C$_2$H, HC$_3$N, CH$_3$CN --  which enables excitation analysis using, e.g.,\ rotational diagrams. In addition, HCN, CN, C$^{17}$O, and C$_2$H lines present resolvable fine or hyperfine structure, which are used to empirically determine line excitation conditions such as density, temperature, and line optical depth, and from there column densities. $c$-C$_3$H$_2$ was also detected in two lines with similar excitation temperatures. The main molecular targets of MAPS --  $^{13}$CO, C$^{18}$O, HCN, C$_2$H, HC$_3$N, and CH$_3$CN -- were observed at both 1 mm (ALMA Band 6) and 3 mm (ALMA Band 3).

Tables \ref{tab:lines-b3} and \ref{tab:lines-b6} list the molecular properties used within the MAPS program.

\begin{deluxetable*}{llccccccccccc}
\tablecaption{Lines targeted at 1~mm (ALMA Band 6)}
\label{tab:lines-b6}
\tablehead{
\colhead{Molecule}  &\colhead{QN}   & \colhead{Rest freq.}    & \colhead{Log$_{10}(A_{\rm ij} / \rm{s}^{-1})$} & \colhead{$g_{\rm u}$} &
\colhead{E$_{\rm u}$}   &\colhead{Cat.$^1$}    &\multicolumn{5}{c}{Detected}\\
\colhead{} &  \colhead{} &\colhead{[GHz]}    &  \colhead{}   & \colhead{} & 
\colhead{[K]}   & \colhead{}&\colhead{IM Lup}   &\colhead{GM Aur}   &\colhead{AS 209}    &\colhead{HD 163296}    &\colhead{MWC 480}
}
\tablecolumns{12}
\startdata
{\bf B6-1}\\    
DCN$^\dagger$   &J=3--2    &217.238538   &    $-$3.3396 & 21 & 20.9   & CDMS  &Y&Y&Y&Y&Y\\
$^{13}$CN  &N=2--1, J=$\frac{3}{2}$--$\frac{1}{2}$, F1=1--0, F=0--1  &217.264639  &$-$3.9547 &1&15.7 &CDMS&N&N&N&N&N\\
& N=2--1, J=$\frac{3}{2}$--$\frac{1}{2}$, F1=1--0, F=1--1 & 217.277680 & $-$3.4953 & 3 & 15.7 & CDMS &N&N&N&N&N\\
& N=2--1, J=$\frac{3}{2}$--$\frac{1}{2}$, F1=2--1, F=2--2 & 217.286804 & $-$3.6541 & 5 & 15.6 & CDMS  &N&N&N&N&N\\
& N=2--1, J=$\frac{3}{2}$--$\frac{1}{2}$, F1=2--1, F=1--1 & 217.290823 & $-$3.6567 & 3 & 15.6 & CDMS  &N&N&N&N&N\\
H$_2$CO    &(J$_{\rm{K}_{\rm{a}}, {\rm{K}_{\rm{c}}}}$)=$3_{03}$--$2_{02}$    &218.222192   &$-$3.5504 &7    &21.0 &JPL &Y&Y&Y&Y&Y\\
C$^{18}$O  &J=2--1   &219.560354 & $-$6.2211 & 5 & 15.8 & LAMDA   &Y&Y&Y&Y&Y\\
$^{13}$CO  &J=2--1   &220.398684  &$-$6.2191 & 5 & 15.9 & LAMDA    &Y&Y&Y&Y&Y\\
CH$_3$CN   &J=12--11, K=3   &220.709017 & $-$3.0624 & 100 & 133.2 & CDMS& N &(Y)$^*$&N&(Y)$^*$&Y\\
&J=12--11, K=2   &220.730261 & $-$3.0465 & 50 & 97.4 & CDMS& N &Y&N&Y&Y\\
&J=12--11, K=1   &220.743011 & $-$3.0372 & 50 & 76.0 & CDMS& (Y)$^*$ &Y&Y&Y&Y\\
&J=12--11, K=0    &220.747262 & $-$3.0342 & 50  & 68.9  & CDMS & N &Y&Y&Y&Y\\
CO &J=2--1  &230.538000   &$-$6.1605 & 5 & 16.6 & LAMDA   &Y&Y&Y&Y&Y\\
N$_2$D$^+$ $^\dagger$ &J=3--2   &231.321828 & $-$3.3586 & 63 & 22.2 & CDMS &Y&(Y)$^*$&Y&Y&Y\\
\hline
{\bf B6-2}\\    
$c$-C$_3$H$_2$ &(J$_{\rm{K}_{\rm{a}}, {\rm{K}_{\rm{c}}}}$)=7$_{07}$--6$_{16}$~$^\ddagger$ &251.314367 & $-$3.0704 & 45  & 50.7 & CDMS &(Y)$^*$&Y&Y&Y&Y\\
&(J$_{\rm{K}_{\rm{a}}, {\rm{K}_{\rm{c}}}}$)=6$_{15}$--5$_{24}$   &251.508708& $-$3.1708 & 13 & 47.5 & CDMS   &N&Y&Y&Y&Y\\
&(J$_{\rm{K}_{\rm{a}}, {\rm{K}_{\rm{c}}}}$)=6$_{25}$--5$_{14}$   &251.527311& $-$3.1706   & 39 & 47.5 & CDMS   &N &Y&N&Y&Y\\
C$_2$H &N=3--2, J=$\frac{7}{2}$--$\frac{5}{2}$, F=4--3   &262.004260   &$-$4.1152 &9&25.2 & CDMS   &Y&Y&Y&Y&Y\\
&N=3--2, J=$\frac{7}{2}$--$\frac{5}{2}$, F=3--2  &262.006482   &$-$4.1321  &7  & 25.2 &CDMS   &Y&Y&Y&Y&Y\\
&N=3--2, J=$\frac{5}{2}$--$\frac{3}{2}$, F=3--2   &262.064986   &$-$4.1521 &7&25.2& CDMS &Y&Y&Y&Y&Y\\
&N=3--2, J=$\frac{5}{2}$--$\frac{3}{2}$, F=2--1    &262.067469   &$-$4.1906 &5& 25.2 &CDMS&Y&Y&Y&Y&Y\\
&N=3--2, J=$\frac{5}{2}$--$\frac{3}{2}$, F=2--2   &262.078935   &$-$5.0619 &5& 25.2 &CDMS&Y&Y&Y&Y&Y\\
HC$_3$N &J=29--28  &263.792308  & $-$2.8349 & 59 & 189.9  & CDMS &N&Y&Y&Y&Y\\
HCN $\dagger$ &J=3--2, F=3--2  &265.886434   & $-$3.1292 & 7.0 & 25.5  & CDMS  &Y&Y&Y&Y&Y\\
&J=3--2, F=3--3 &265.884891 &$-$4.0322 & 7.0 & 25.5  & CDMS & Y & Y & Y & Y & Y\\
&J=3--2, F=2--2 &265.888522 & $-$3.8861 & 5.0 & 25.5  & CDMS & Y & Y & Y & Y & Y\\
\enddata
\tablenotetext{}{$^1$ Spectroscopic data are compiled from the CDMS \citep{Muller01, Muller05,Endres16}; JPL \citep{Pickett98}; and LAMBDA \citep{Schoier05} catalogues.}
\tablenotetext{}{$\dagger$ See \citet{Cataldi21} for the full list of hyperfine components.}
\tablenotetext{}{$\ddagger$ Ortho/para blend; see \citet{Ilee21} for the full line list.}
\tablenotetext{}{* Tentatively detected with a 3-5$\sigma$ significance \citep{Cataldi21,Ilee21}.}
\end{deluxetable*}

\section{Observational Details, Calibration and Imaging} \label{sec:obs}

\subsection{Observations}

MAPS (2018.1.01055.L) was executed between October 2018 and September 2019 (with some remaining executions scheduled for 2021). The short baseline executions were observed between October 2018 and April 2019, while all long baseline executions were taken in August and September 2019. In total 80 executions have been carried out, most of which consisted of $\sim$45~min on target and 20--45~min of calibration, with longer calibration times required for the longer baseline observations. Tables \ref{tab:obs-b3} and \ref{tab:obs-b6} in Appendix \ref{app:obs} list the observational details including observing dates, number of antennas, integration times, baselines, approximate angular resolution, maximum recoverable scale, phase and flux calibrators for each execution. Note that GM Aur and MWC 480 always shared tracks, and their listed integration times were exactly split between the two sources in each execution.

\begin{deluxetable*}{llccccccc}
\tablecaption{Correlator set-ups in ALMA Bands 3 and 6}
\label{tab:corr}
\tablehead{
\colhead{Set-up}   &\colhead{Center Freq.}  &\colhead{Line Targets\tablenotemark{$^a$} }   & \colhead{Resolution}    & \colhead{Bandwidth}\\
 \colhead{}   & \colhead{[GHz]} &  \colhead{} &\colhead{[km~s$^{-1}$]}    & \colhead{[MHz]} 
}
\startdata
B3-1    &86.054966  &HC$^{15}$N J=1--0    &0.492  &58.59\\
&86.339918  &H$^{13}$CN J=1--0    &0.490  &58.59\\
&86.754288  &H$^{13}$CO$^+$ J=1--0    &0.488  &58.59\\
&87.310000  &C$_2$H N=1--0    &0.485  &58.59\\
&88.631601  &HCN J=1--0   &0.239  &58.59\\
&89.188526  &HCO$^+$ J=1--0   &0.237  &58.59\\
&99.000000  &Continuum band   &3.419  &1875\\
&100.076392  &HC$_3$N J=11--10   &0.211  &58.59\\
&101.332991  &H$_2$CO (J$_{\rm{K}_{\rm{a}}, {\rm{K}_{\rm{c}}}}$)=$6_{15}$--$6_{16}$   &0.209  &58.59\\
\hline
B3-2    &97.980953  &CS J=2--1    &0.216  &117.19\\
&100.000000  &Continuum band    &3.385  &1875\\
&109.782176  &C$^{18}$O J=1--0    &0.385  &58.59\\
&110.201354  &$^{13}$CO J=1--0    &0.384  &58.59\\
&110.381346  &CH$_3$CN J=6--5, K=0--5    &0.383  &117.19\\
&112.359278  &C$^{17}$O J=1--0    &0.188  &58.59\\
&113.499644  &CN N=1--0    &0.186  &58.59\\
\hline
B6-1    &217.238530  &DCN J=3--2, $^{13}$CN N=2--1 &0.195  &117.19\\
&218.222192  &H$_2$CO (J$_{\rm{K}_{\rm{a}}, {\rm{K}_{\rm{c}}}}$)=$3_{03}$--$2_{02}$  &0.194  &117.19\\
&219.560358  &C$^{18}$O $2-1$  &0.193  &58.59\\
&220.398684  &$^{13}$CO $2-1$  &0.192  &58.59\\
&220.709099  &CH$_3$CN  J=12--11, K=2--4   &0.192  &58.59\\
&220.743097  &CH$_3$CN J=12--11, K=0--2    &0.192  &58.59\\
&230.538000  &CO J=2--1  &0.092  &58.59\\
&231.321828  &N$_2$D$^+$ J=3--2  &0.091  &58.59\\
&234.000000  &Continuum band  &1.446  &1875\\
\hline
B6-2    &249.000000  &Continuum band  &1.359  &1875\\
&251.314337  &$c$-C$_3$H$_2$ (J$_{\rm{K}_{\rm{a}}, {\rm{K}_{\rm{c}}}}$)=7$_{07}$--6$_{16}$  &0.168  &117.19\\
&251.527302  &$c$-C$_3$H$_2$  (J$_{\rm{K}_{\rm{a}}, {\rm{K}_{\rm{c}}}}$)=6$_{25}$--5$_{14}$  &0.168  &117.19\\
&262.040000  &C$_2$H N=3--2   &0.161  &117.19\\
&263.792308 &HC$_3$N J=29--28   &0.160  &117.19\\
&265.886431 &HCN J=3--2  &0.159  &234.38\\
\enddata
\tablenotetext{}{$a$ A full listing of line properties is found in Tables \ref{tab:lines-b3} and \ref{tab:lines-b6}.}
\end{deluxetable*}

A key aspect of the MAPS project is the correlator set-up, which is outlined in Table \ref{tab:corr} for the two ALMA Band 3 (B3-1 and B3-2) and Band 6 (B6-1 and B6-2) observing modes. Each correlator set-up consist of 6--9 spectral windows (SPWs), one of which is a continuum band with a coarse (1.4--3.4 km s$^{-1}$) velocity resolution. The latter is needed to self-calibrate (see next section). The remaining SPWs are designed to have a velocity resolution of 0.19--0.49 km~s$^{-1}$ in B3-1 and B3-2, and 0.09--0.20 km~s$^{-1}$ in B6-1 and B6-2, and no online spectral binning was applied. These velocity resolutions are well below typical disk-averaged line widths of a few km~s$^{-1}$, but comparable to the intrinsic line widths of $<$0.5 km~s$^{-1}$. The highest resolution of 0.09 km/s is used for the CO $2-1$ line, to enable kinematic studies. Other expected strong lines were observed at $<$0.2 km~s$^{-1}$ and weaker lines at coarser resolution unless paired with a strong line. In each case, we used the maximum resolution possible for the chosen bandwidth. For the delivered data products the velocity resolutions have been coarsened to achieve more uniformity between spectral-line cubes as described in \citet{Czekala21}. 

\subsection{Calibration Strategy}

Our calibration strategy closely follows the one developed for the DSHARP ALMA Large Program, which is described in detail by \citet{Andrews18}. We summarize the strategy here and provide more extensive descriptions of any procedures that deviated from the DSHARP script.

All data were initially calibrated by ALMA staff using the ALMA calibration pipeline. The short and long baseline data were calibrated separately, and because some of the short baseline data were observed a year before the last long baseline observations were completed, not all data were calibrated with the same pipeline version. CASA 5.4 was used for the observations carried out in 2018, and CASA 5.6 for the observations carried out in 2019.
The observations typically have $T_{\rm sys} \sim$50--80 K, and the rms phase variations after Water Vapor Radiometer (WVR) corrections were generally within a range of $\sim 15^{\circ}$--$50^{\circ}$.

All data were self-calibrated. Prior to self-calibration, we carried out a number of data processing steps. Unless otherwise noted, we used CASA 5.6 for these steps as well as for self-calibration. We used CASA in parallel mode for all tasks where it is supported, but due to a bug in CASA 5.6, we had to switch to CASA 5.4 when running the \texttt{gaincal} and \texttt{virtualconcat} tasks. 

As a first step towards self-calibration, we created pseudo-continuum visibilities by flagging the line emission in each spectral window. Except for the case of CO 2--1 towards HD~163296, we flagged channels with velocities between $-10$ and 20 km~s$^{-1}$ (all sources have systemic velocities between 4.5 and 5.8 km~s$^{-1}$). In the SPW containing CO 2--1 towards HD~163296, we extended the flagged region to include its disk wind \citep{Klaassen13}. Second, we averaged-down the pseudo continuum data into 125~MHz channels, and imaged each execution block. We identified the continuum peak emission, and used it to align the different executions to a common phase center. 

We self-calibrated the aligned data starting with short-spacings (short baseline data). The self-calibrated short-spacing data were then concatenated with the long-baseline data, and the combined visibilities were self-calibrated together. In the self-calibration of the Band 3 short-spacing data, and in the self-calibration of the combined data of both settings, we combined SPWs and scans to improve the signal-to-noise ratio (SNR). That was not necessary for the Band 6 short-spacing data due to its higher continuum brightness. We performed between one and six iterations of phase-only self-calibrations, and then tried one amplitude self-calibration. We typically started with a solution interval of 900s, and then stepped down to 360, 180, 60, 30, and 18s in each iteration. The reference antenna was chosen from the log  based on its data quality and position in the array. 

We applied the calibration table to spectrally averaged visibilities in each iteration, and  imaged the data with the \tclean\ task using a Briggs robust parameter of 0.5 and an elliptical mask. The mask dimensions and position angle were selected based on the inclination and position angle of each disk. The resulting peak intensity and the image SNR were used to determine whether to proceed with the next iteration or not: if the peak intensity increased and the signal-to-noise improved by a factor larger than 5\%, we proceeded with the next iteration choosing a shorter solution interval. After reaching the stopping criteria for phase self-calibration, we attempted one round of amplitude self-calibration and used the results if the peak SNR increased and the image quality visually improved. Finally, the resulting calibration tables were applied to the aligned, but unflagged and spectrally non-averaged visibilities. We used the task \texttt{uvcontsub} to subtract the continuum, providing a .contsub measurement set for each disk/setting. These data products as well as continuum+line measurement sets are available for download from the ALMA Archive via \url{https://almascience.org/alma-data/lp/maps}.

The improvements in the continuum images following self-calibration were substantial for the Band 6 settings, and smaller for the Band 3 settings.  The CLEAN beam was typically around 0\farcs1 in band 6 and 0\farcs3 in Band 3, with an RMS of $10-20$ $\mu$Jy/beam. The continuum image properties are discussed in \citet{Sierra21} and we refer the reader to this paper for more details on the MAPS continuum data.

\subsection{Summary of Imaging Strategy and Products}

The scope of MAPS and its range of line targets observed at high resolution required substantial development work to produce accurate and aesthetically pleasing imaging products. The MAPS imaging strategy, and its motivations and verification are described in detail in \citet{Czekala21}. Here we summarize key aspects of the process. Prior to imaging, the data were split into individual measurement sets for each targeted line, and regridded onto a common velocity grid using the task \texttt{cvel2} in CASA. A velocity spacing of 0.2 km~s$^{-1}$ was used for the Band 6 data and 0.5 km~s$^{-1}$ for the Band 3 data. Channels were centered on the systemic velocity of each disk (see Table \ref{tab:sources}) and extended out to 1.2x the empirically determined extent of the $^{12}$CO 2--1 emission for each disk. This procedure was repeated for both the continuum subtracted data and the data still containing continuum emission. The MAPS images were produced using the task \tclean\ in CASA. Unless stated otherwise, we built \clean\ masks assuming Keplerian rotation, and the stellar masses, disk inclinations and position angles listed in Table \ref{tab:sources}. The outer radii of these masks were set to generously incorporate all emission from $^{13}$CO 2--1, which had the most extended emission structure except $^{12}$CO 2--1, towards each disk. Because $^{12}$CO 2--1 displays non-Keplerian structure in some of the disks, we used manual, hand-drawn masks for this line. Emission was cleaned down to 4$\times$ RMS, estimated in a line free region. See \S5 of \citet{Czekala21} for more details on this decision.

The Band 6 and Band 3 spectral-line cubes were initially cleaned to achieve high spatial resolution, while simultaneously achieving good image fidelity. Using a robust parameter of 0.5, we achieved a beam minor and major axes full-width at half-maximum (FWHM) of 0\farcs09--0\farcs13 and 0\farcs11--0\farcs17, respectively in Band 6 and 0\farcs21--0\farcs32 and 0\farcs26--0\farcs47, respectively in Band 3.  This corresponds to spatial resolutions of 10--20~au in Band 6, and 20--50~au in Band 3. We found that applying a small taper to the Band 6 images substantially improved the sensitivity to low surface brightness features and resulted in smoother images, while minimally decreasing the resolution. For our fiducial images, we applied tapers that resulted in circularized  0\farcs15 beam for the Band 6 data. We also re-imaged the Band 3 data with a circularized 0\farcs3 beam, and these are the fiducial Band 3 images. Unless stated otherwise, these fiducial image cubes form the basis of the analysis within the MAPS collaboration.  In addition, the Band 6 line image cubes were also tapered to achieve 
 beams of 0\farcs2 and 0\farcs3, respectively. The latter facilitated imaging of the weakest lines, as well as comparisons between Band 3 and Band 6 lines. To increase the SNR for the weakest Band 3 lines we also tapered the Band 3 images to a resolution of 0\farcs5. Table \ref{tab:im-products} summarizes the available image cube resolutions.
 
 \begin{deluxetable}{cccc}
\tablecaption{Overview of spectral image cube spatial resolutions}
\label{tab:im-products}
\tablehead{
\colhead{B3-1}   &\colhead{B3-2}  &\colhead{B6-1}   & \colhead{B6-2}    
}
\startdata
0\farcs22--0\farcs32$^\dagger$& 0\farcs21--0\farcs28$^\dagger$& 0\farcs10--0\farcs13$^\dagger$& 0\farcs09--0\farcs11$^\dagger$ \\
& & {\bf 0\farcs15$^\star$} &{\bf 0\farcs15$^\star$}\\
& & 0\farcs2 &0\farcs2\\
{\bf 0\farcs3$^\star$} &{\bf 0\farcs3$^\star$}  & 0\farcs3 &0\farcs3\\
0\farcs5 &0\farcs5  \\
\enddata
\tablenotetext{}{$^\dagger$ Resolution range refers to the minor beam axis when using robust=0.5.}
\tablenotetext{}{$^\star$ These are the fiducial MAPS imaging products.}
\end{deluxetable}

\begin{figure*}[ht!]
\centering
\includegraphics[scale=0.65]{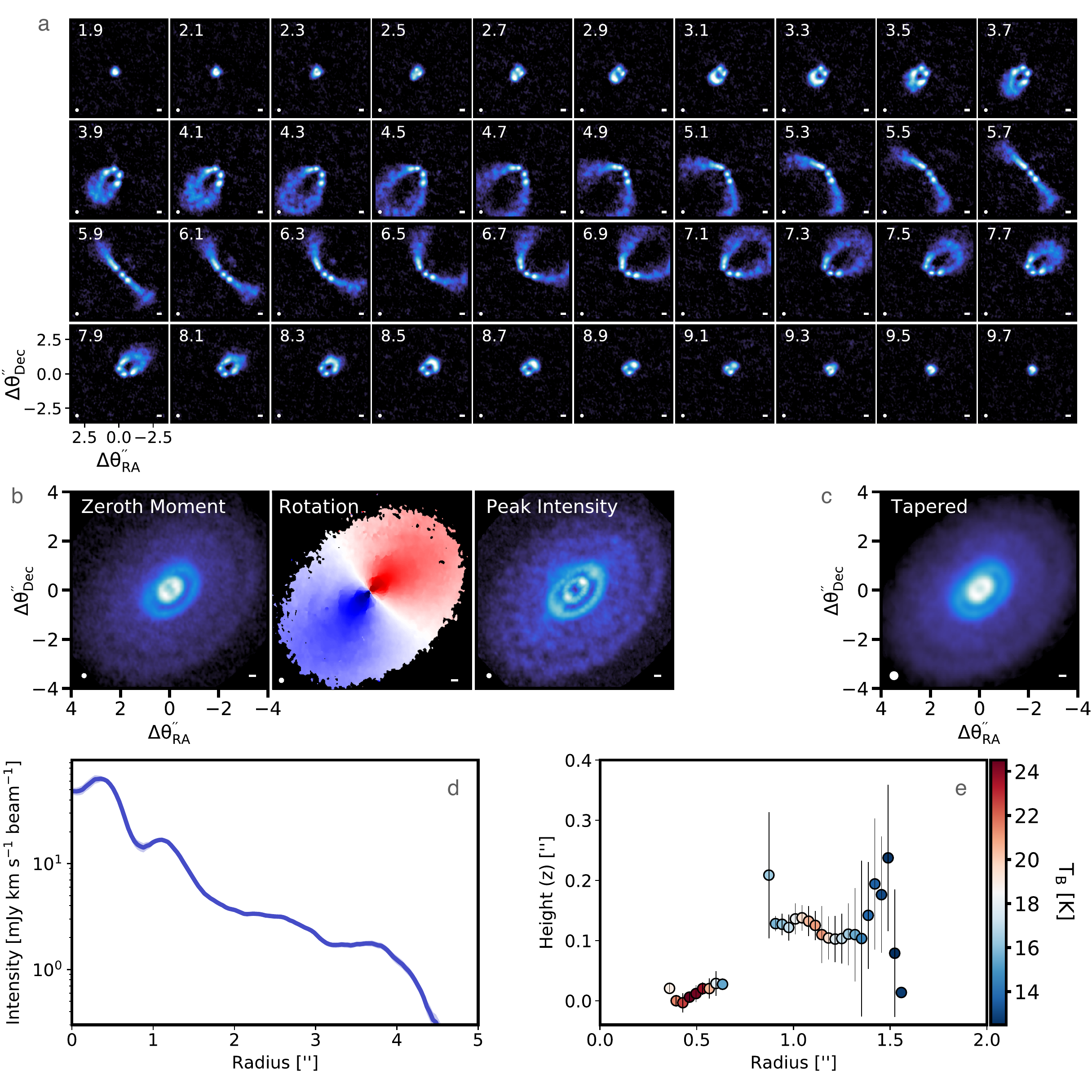}
\caption{Illustration of MAPS imaging and image-derived products using the HCN 3--2 line towards the HD 163296 disk as an example. a) 40 channels of the fiducial-resolution (0\farcs15) spectral-line cube with LSRK velocities (km s$^{-1}$)  noted in the upper right corners. b) Zeroth moment, rotation, and peak intensity maps generated from the fiducial spectral-line cube shown in a). In the rotation map, darker colors correspond to higher absolute velocities. c) The zeroth moment map generated from a spectral line cube with 0\farcs3 taper to enhance sensitivity to weak, large-scale emission. d) The radial profile of the HCN fiducial zeroth moment map extracted from a 15$^\circ$ wedge along the disk major axis. e) the z vs r distribution of HCN emission, where the color encodes brightness temperature between 12 (blue) and 25 (red) K. The channel maps (a), the zeroth moment and the peak temperature brightness ($T_{\rm B}$) map (in b) are plotted with a power law color stretch normalized to peak flux in each panel to visualize both weak and strong features. The synthesized beam is shown in the bottom-left corner in each panel, and a 20 au scale bar in the bottom-right corners. \label{fig:imaging}}
\end{figure*}

After generating the image cubes, image residuals were scaled by a factor $\epsilon$, equal to the ratio of the CLEAN beam and dirty beam effective areas, to account for the effects of non-Gaussian beams  (resulting from joint configuration imaging) on the image quality, and to  provide a realistic image signal-to-noise ratio \citep{Jorsater95}. Throughout MAPS, we refer to this as the JvM correction, and it is needed because the dirty beam was highly non-Gaussian when combining the short and long baseline configurations \citep[see][for further details]{Czekala21}. The applied JvM corrections to the image residuals are listed in Table \ref{tab:wide_table_rms} in Appendix \ref{app:rms}  for our fiducial image line cubes together with the measured RMS (after application of the JvM correction). Figure \ref{fig:imaging}a visualizes 40 of the channels belonging to the HD~163296 fiducial HCN 3--2 spectral-image cube, i.e., a channel map. The Keplerian rotation of the disk is visible, as is some radial substructure.

\subsection{Value-Added Data Products (VADPs)}

In addition to spectral image cubes, we have generated a range of higher-level image products and radial profiles, referred to collectively as Value-Added Data Products (VADPs). The process through which these were generated and their detailed descriptions can be found in \citet{Law21_radprof}. Figure \ref{fig:imaging} shows some examples of these image products. In particular Figure \ref{fig:imaging}b shows the fiducial (0\farcs15) zeroth moment (integrated flux), rotation map (velocity field), and peak intensity (spectral line maximum intensity) maps of the HD~163296 HCN 3--2 line. The zeroth moment map clearly shows two small-scale rings as well as one or two diffuse rings in the outer disk. The rotation map shows a characteristic Keplerian velocity field. Fig.\ \ref{fig:imaging}c shows the zeroth moment map generated from the low resolution 0\farcs3 spectral-image cube. Compared to the fiducial zeroth moment map, the inner rings are less sharp, while the outer disk structure is more clearly visible, demonstrating the utility of imaging the same line and disk with different tapers. Note that most moment maps shown in MAPS papers have been optimized to visualize substructure and best show the quality of the data using `hybrid' masks, as described in \citet{Law21_radprof}. For any quantitative science, we strongly recommend using zeroth moment maps generated without the hybrid masks, or to use line image cubes or radial profiles.

The top level image-derived products are radial profiles and emission surfaces. The radial profiles are generated from zeroth moment maps constructed using only Keplerian masks (i.e., not from the `hybrid' zeroth moment maps described above) and take into account the derived molecular emission heights when available \citep{Law21_surf}. For lines with no empirical constraints on their emission surfaces, we assume that the emission originates from $z/r=0$, i.e.,\ the midplane \citep{Law21_radprof}.  In each case the emission is deprojected using the disk inclinations and position angles listed in Table \ref{tab:sources} and python package \texttt{GoFish} \citep{Teague19b}. The emission is then averaged over a range of wedge sizes along the major axis as well as over the entire azimuth to generate a series of radial profiles for each line \citep{Law21_radprof}. Figure \ref{fig:imaging}d shows the radial profile extracted from the fiducial spectral image cube of HCN in the HD~163296 disk using a narrow wedge. Similar to the moment maps, the HCN radial profiles show two rings in the inner disk, a plateau or broad ring between 2\arcsec and 3\arcsec, and a faint ring between 3\arcsec and 4\arcsec, but with a clearer view of the relative fluxes of the different components compared to the moment maps. 

Figure \ref{fig:imaging}e shows an example of an emission surface, color-coded by brightness temperature.  Following, e.g.,  \citet{Rosenfeld13}, \citet{Isella18} and \citet{Pinte18}, emission surfaces are extracted from  the  image cubes for molecular lines where the front and back sides of the disks can be resolved, resulting in a clearly observed emission asymmetry relative to the major axis. Practically, we used the \texttt{disksurf}\footnote{\url{https://github.com/richteague/disksurf}} python package which has implemented this method together with different data filtering and smoothing capabilities \citep{Law21_surf}.  This extraction method requires strong emission and high spatial resolution, and is only carried out for the brightest emission lines, i.e., the CO 2--1 isotopologues, HCN 3--2, and C$_2$H 3--2.

The described VADPs are generated for all lines, disks, and spatial resolutions.  

\subsection{Image Repository Description}

The image cubes and value added data products are available for download via the ALMA archive, which can be accessed directly at \url{https://almascience.nrao.edu/asax/} or through the MAPS ALMA landing page at \url{ https://almascience.nrao.edu/alma-data/lp/maps}. An interactive browser for this repository is available on the MAPS project homepage at \url{http://www.alma-maps.info}. The spectral line cubes and associated products are described in detail in \citet{Czekala21}, and the VADPs in \citet{Law21_radprof} and \citet{Law21_surf}.

The data are structured hierarchically by [disk] $\rightarrow$ [spectral setting] $\rightarrow$ [molecule] $\rightarrow$ [line], and then each line has available the relevant windowed measurement set, image cubes (at each available resolution), corresponding VADPs, and the scripts needed to replicate each stage of the imaging and VADP process for that line. The continuum subtracted and non-continuum subtracted data and products are separated at the spectral setting level. Both primary-beam corrected and non-primary beam corrected images are available, as are the CLEAN masks used for each image. We  provide both the JvM corrected images (using the noted values of $\epsilon$) and uncorrected images, but  recommend that JvM corrected images are used to do science.

Images are named \begin{verbatim}
[disk]\_[molecule]\_[spectral setup]\
_[line identifier  (hyperfine/k-ladder
 if needed)].[imaging resolution].clean.
[JvMcorr].image.[pbcor].fits, 
\end{verbatim} where [JvMcorr] and [pbcor] are present if those corrections were made. Continuum subtraction is noted in the imaging resolution flag (e.g.,\ `0.3arcsec\_wcont'). CLEAN masks are named similarly, but with a .mask.fits extension. Radial profiles are named: [disk]\_[molecule]\_[spectral setup]\_[line identifier (hyperfine/k-ladder, if needed)].[imaging resolution].[wedge size]\_radialProfile.txt, where [wedge size] denotes the angular extent of the extraction wedge used (e.g.,\ `15deg') or `azimuthal' for the full azimuthally averaged profile.

\subsection{Line detections}

The vast majority of the targeted lines were detected, resulting in $\sim$200 line detections across the five disks. Tables \ref{tab:lines-b3} and \ref{tab:lines-b6} note three cases: detections, tentative detections, and non-detections.  We consider a line detected when its integrated emission, stacked spectra \citep{Teague19b}, and/or peak matched filter response is $>$5~sigma \citep{Loomis18a}. A small set of lines were not clearly detected in the image cubes, and had a peak matched filter response of 3--5$\sigma$. We consider these tentatively detected and they are labeled (Y) in the tables. 

We detect all targeted CO isotopologue lines (except for C$^{17}$O), all targeted strong HCN, C$_2$H, DCN and CN line components, as well as HCO$^+$ 1--0, CS 2--1, and H$_2$CO $3_{03}$--$2_{02}$ lines towards all disks. Lines from H$^{13}$CO$^+$ 1--0, HC$_3$N 11--10 and 29--28, C$^{17}$O 1--0, N$_2$D$^+$ 3--2, and all three $c$-C$_3$H$_2$ lines, as well as the strongest CH$_3$CN lines were  detected towards 3 or 4 disks each and often tentatively detected towards an additional disk. Some of the higher energy levels and therefore weaker CH$_3$CN lines were detected towards 1--2 disks, and two of the targeted CH$_3$CN lines were undetected towards all disks. A few additional lines, including H$^{13}$CN 1--0, HC$^{15}$N 1--0, $^{13}$CN 3--2, and H$_2$CO $\rm 6_{15}-6_{16}$, were also not detected towards any disks, except for one tentative H$^{13}$CN detection towards GM Aur. 

Appendix \ref{app:lines} reports the approximate disk-integrated line intensities (or upper limits) for all disks and lines. Note that these values were generated using an automated script that calculates the flux within the Keplerian CLEAN mask described in \citet{Czekala21}. We hope that this overview is useful for, e.g.,\ observation preparation, but do not recommend that it is used for any quantitative analysis. For the latter, we refer the reader to the more precise disk-integrated values and radial disk profiles presented in the individual MAPS papers \citep{Zhang21,Guzman21,Cataldi21,Ilee21,Bergner21,LeGal21, Aikawa21}.
Still, these disk integrated fluxes do provide some useful measures of the range of disk integrated line fluxes across the sample. The CO 2--1 integrated fluxes are $\sim$7.8--45 Jy km s$^{-1}$, while the $^{13}$CO 2--1 and C$^{18}$O 2--1  line fluxes range between $\sim$2.3--16 and $\sim$0.54--5.8 Jy km s$^{-1}$, respectively. The corresponding 1--0 lines are  weaker at  $\sim$0.38--3.6 and  $\sim$0.073--0.98 Jy km s$^{-1}$ for $^{13}$CO 1--0 and C$^{18}$O 1--0, respectively. Among the other Band 6 lines, HCN and C$_2$H 3--2 line fluxes are the highest ($\sim$1.8--7.5 and 0.57--3.5 Jy km s$^{-1}$, respectively), while among the Band 3 lines, the HCO$^+$ 1--0 line is the strongest non-CO line  at $\sim$0.20--1.1 Jy km s$^{-1}$.

The line emitting areas, characterized as the disk radius that encloses 90\% of the flux, are presented in detail by \citet{Law21_radprof}. Here we simply note a few statistics. The line emission disk radii range from $\sim$50 to 500~au. The largest radii are seen for CO 2--1, HCO$^+$ 1--0, CS 1--0, CN 1--0, and H$_2$CO $3_{03}$--$2_{02}$, which all have radii of 300-500~au in a majority of disks. By contrast, CH$_3$CN and HC$_3$N present small line emission radii, between 50 and 100~au. For individual lines, we do see up to a factor of five differences in emission areas between disks.

\begin{figure*}[ht!]
\centering
\includegraphics[scale=0.7]{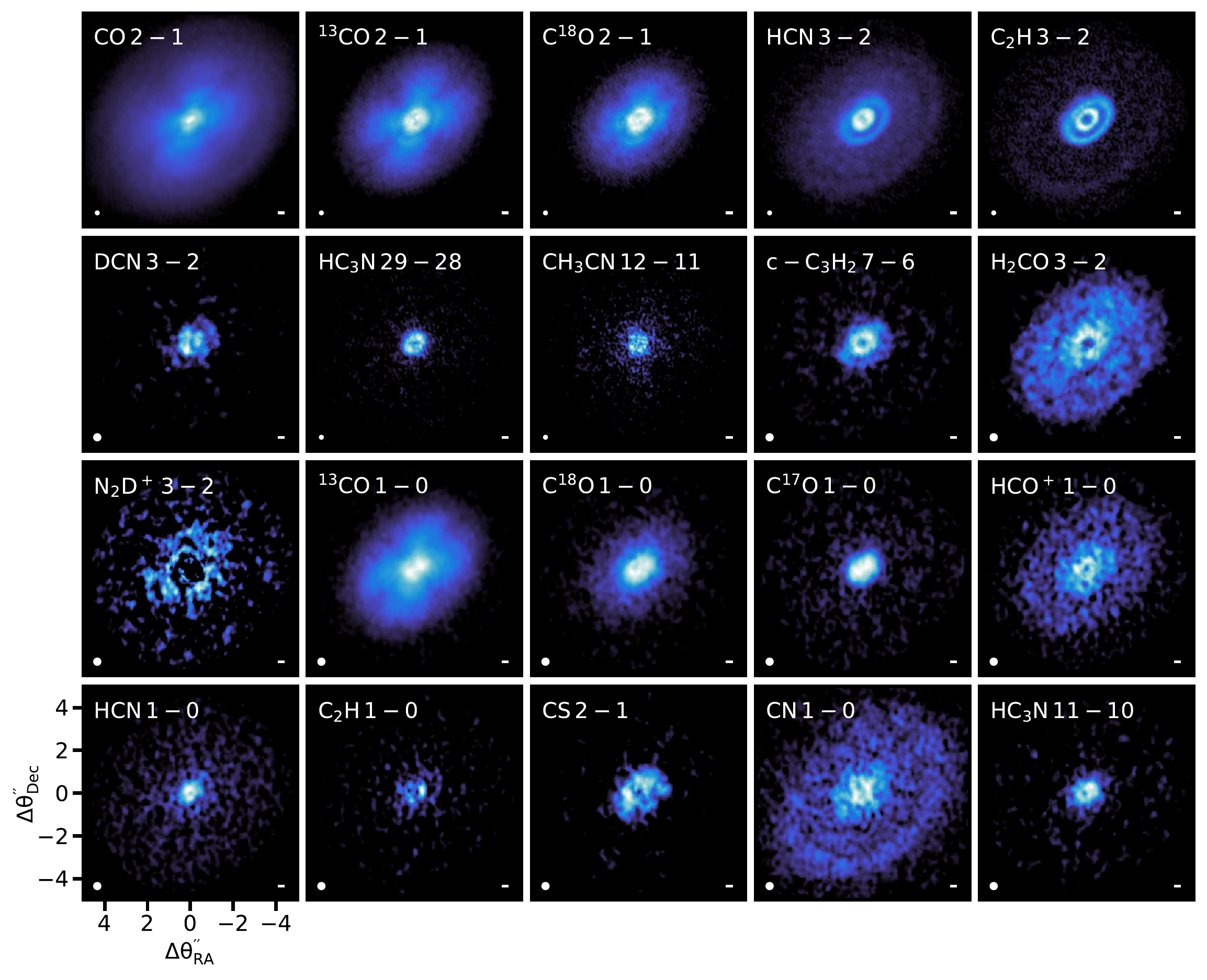}
\caption{20 molecular faces of the HD~163296 disk. These comprise a representative, but non-exhaustive, set of zeroth moment maps towards HD~163296. For those lines with multiple hyperfine lines, the brightest component is shown. The images are shown with a 5--10 mJy km$^{-1}$ clip to minimize the visual contribution of noise and with a powerlaw color stretch to emphasize the outer disk fainter structures. The color scale is normalized to the peak intensity in each panel. The synthesized beam (0\farcs15 or 0\farcs3) is indicated in the bottom left corner, and a 20~au scale bar is shown in the bottom right corner. \label{fig:hd-gallery}}
\end{figure*}

\section{Overview of MAPS Results}\label{sec:res}

The purpose of this section is to provide an overview of the results from the  18 MAPS papers following this one and \citet{Czekala21}, and to provide some guidance of how the individual results fit together. The MAPS papers span a range of topics, and are here roughly divided into papers focusing on 1) an empirical analysis of the radial and vertical distribution of line emission (MAPS III--IV), 2) retrieval of molecular column densities and abundances, and their interpretation (MAPS V--XIII), 3) disk dust and gas properties, including non-axisymmetric structures and dynamics (XIV--XX). When citing any of these results, the reader is requested to reference the original papers:  MAPS II-XX (Table \ref{tab:papers}).
 
 \begin{deluxetable*}{lll}
\tablecaption{MAPS Papers}
\label{tab:papers}
\tablehead{
\colhead{MAPS}&\colhead{Reference}   &\colhead{Title} 
}
\startdata
I   &This paper    &Program Overview and Highlights\\
II  &\citet{Czekala21}    &CLEAN Strategies for Synthesizing Images of Molecular Line Emission in\\ && Protoplanetary Disks\\
III &\citet{Law21_radprof}    &Characteristics of Radial Chemical Substructures\\
IV  &\citet{Law21_surf}    &Emission Surfaces and Vertical Distribution of Molecules\\
V   &\citet{Zhang21}  &CO Gas Distributions\\
VI  &\citet{Guzman21} &Distribution of the Small Organics HCN, C$_2$H, and H$_2$CO\\
VII &\citet{Bosman21_VII}  &Sub-stellar O/H and C/H and super-stellar C/O in Planet Feeding Gas\\
VIII    &\citet{Alarcon21}  &Gap Chemistry in AS 209 -- Gas Depletion or Chemical processing?\\
IX  &\citet{Ilee21} & Distribution and Properties of the Large Organic Molecules HC$_3$N, CH$_3$CN, and\\&& $c$-C$_3$H$_2$\\
X   &\citet{Cataldi21}  &Studying deuteration at high angular resolution towards protoplanetary disks\\
XI  &\citet{Bergner21}  &CN and HCN as Tracers of Photochemistry in Disks\\
XII &\citet{LeGal21}    &Inferring the C/O and S/H Ratios in Protoplanetary Disks with Sulfur Molecules\\
XIII    &\citet{Aikawa21}   &HCO$^+$ and Disk Ionization Structure\\
XIV &\citet{Sierra21}   &Revealing Disk Substructures in Multi-wavelength Continuum Emission\\
XV  &\citet{Bosman21_XV}  &Tracing Proto-planetary Disk Structure within 20 au\\
XVI &\citet{Booth21b}    &Characterising the Impact of the
Molecular Wind on the Evolution of the \\&&HD 163296 System\\
XVII    &\citet{Calahan21} & Determining the 2D Thermal Structure of the HD 163296 Disk\\
XVIII   &\citet{Teague21}  &Kinematic Substructures in the Disks of HD 163296 and MWC 480\\
XIX &\citet{Huang21}   &Spiral Arms, a Tail, and Diffuse Structures Traced by CO around the GM Aur Disk\\
XX  &\citet{Schwarz21}  &The Massive Disk Around GM Aurigae\\
\enddata
\end{deluxetable*}

\subsection{Radial and vertical distributions of molecular emission}

Zeroth moment maps and radial line emission profiles, and vertical line emission surfaces  are presented in \citet{Law21_radprof} and \citet{Law21_surf}, respectively, herein referred to as {\bf MAPS III} and {\bf MAPS IV} (see also \S3.4 of this paper for information on how moment maps, emission surfaces, and radial profiles were extracted). Here we present some of the high-level results and illustrative examples.

\begin{figure*}[ht!]
\centering
\includegraphics[scale=0.24]{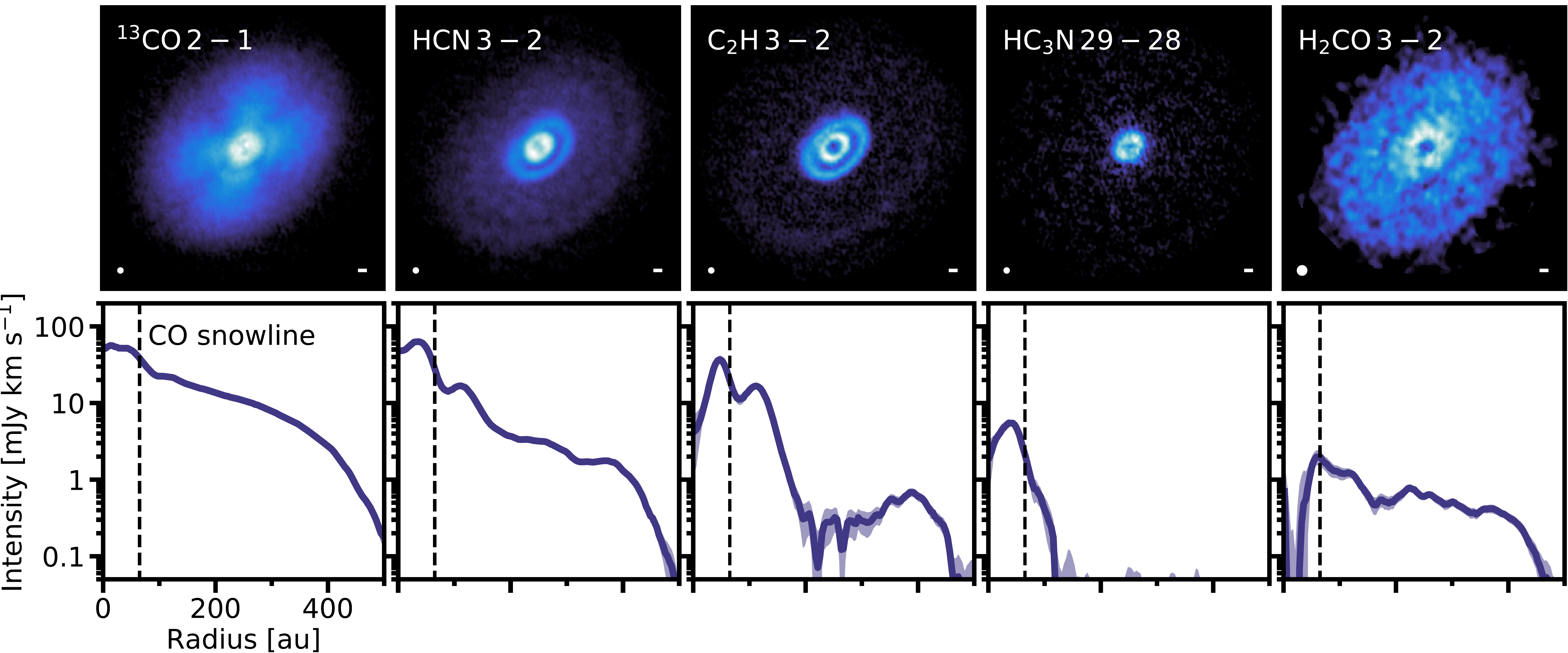}
\caption{Comparisons of emission morphologies between different molecules within the HD 163296 disk. Top row: Zeroth moment maps are shown with a 5 mJy km$^{-1}$ clip, and with a powerlaw color stretch to emphasize the outer disk fainter structures. The color scale is normalized to the peak intensity in each panel. For those lines with multiple hyperfine lines, the brightest component is shown. The synthesized beam is indicated in the bottom left corner, and a 20~au scalebar in the bottom right corners. Bottom: Deprojected and azimuthally-averaged radial normalized intensity profiles. The dashed lines mark the estimated CO snowline location.  \label{fig:comp-hd}}
\end{figure*}

Fig.\ \ref{fig:hd-gallery} shows a  gallery of zeroth moment images towards HD~163296. The amount of spatial diversity is astonishing -- HD~163296 almost looks like a different disk in each molecular line. The degree of diversity is highlighted in Figure \ref{fig:comp-hd} (upper panels), which shows the zeroth moment maps together with radial profiles of five of the brighter molecular lines. Among these five lines, we see one centrally peaked disk with some superimposed shallow rings and gaps ($^{13}$CO), 3--4 nested rings (HCN and C$_2$H), one compact ring (HC$_3$N), and two very broad rings (H$_2$CO). Different lines clearly respond differently to the local disk environment, which suggests that there are multiple causes of line substructures. Indeed, MAPS III finds that while many of the chemical substructures coincide with dust substructures, pebble disk edges, or inferred snowline locations, others do not. This implies that many chemical substructures observed in disks are caused in whole or in part by additional processes, likely including UV photon fluxes, ionization, radially varying elemental ratios, and gas and dust temperatures. 

\begin{figure*}[ht!]
\centering
\includegraphics[scale=0.7]{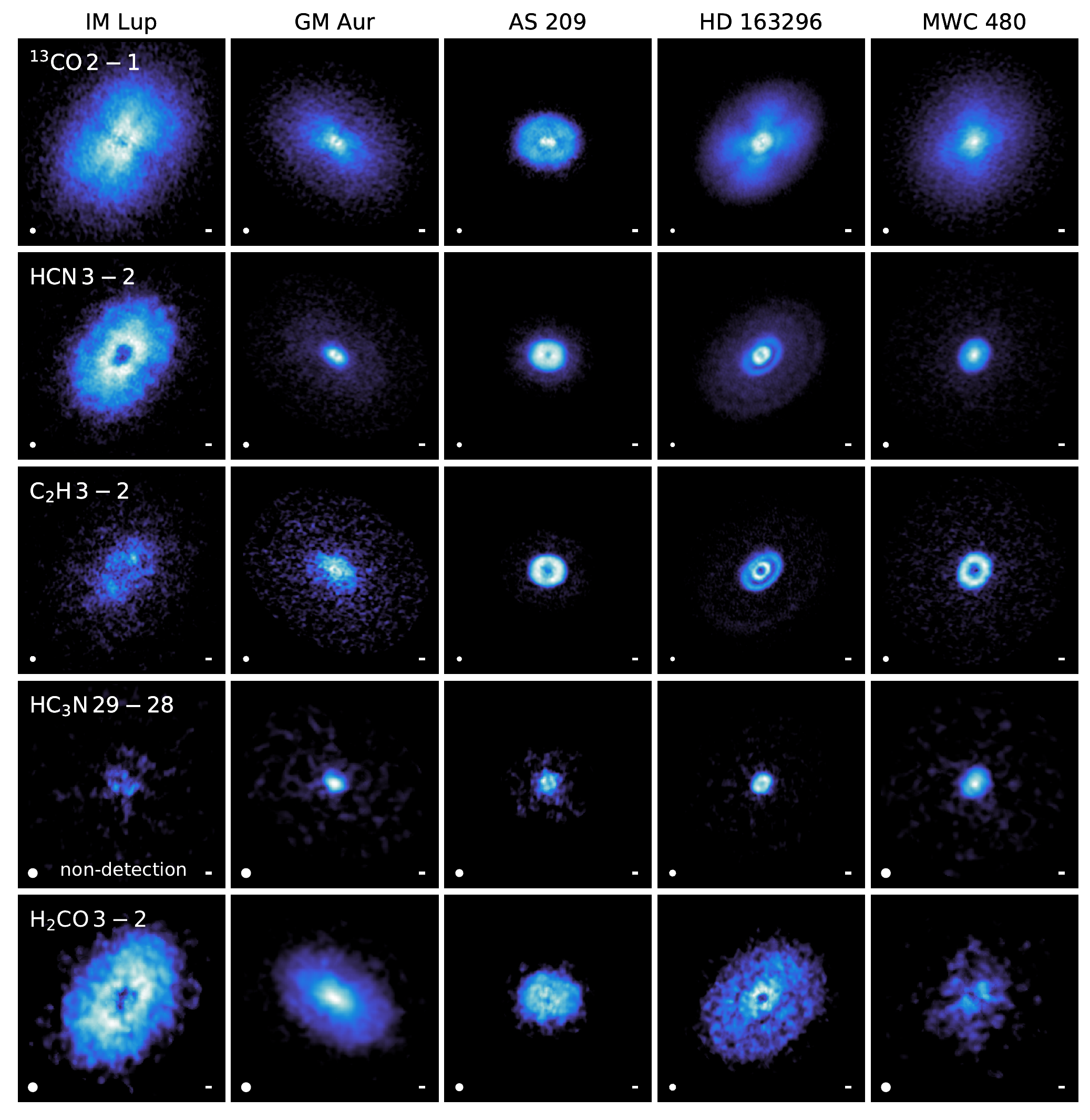}
\caption{Zeroth moment maps of $^{13}$CO 2--1, HCN 3--2, C$_2$H 3--2, HC$_3$N 29--28, and H$_2$CO 3--2 towards (left to right) IM Lup, GM Aur, AS 209, HD 163296, and MWC 480. For those lines with multiple hyperfine lines, the brightest component is shown. The angular box sizes are different for the different disks and have been chosen to show the disks on the same physical scales. The small horizontal bar in the lower right corner corresponds to 20~au along the disk major axis. The images are shown with a 10 mJy km$^{-1}$ clip, and a powerlaw color stretch to emphasize the outer disk fainter structures. The color scale is normalized to the peak intensity in each panel. \label{fig:gallery}}
\end{figure*}

\begin{deluxetable*}{lccccccc}
\tablecaption{Chemical substructure statistics\label{tab:ring_gap_stats}}
\tablewidth{0pt}
\tablehead{
\colhead{Source} &\multicolumn{3}{c}{Rings} & \colhead{} &\multicolumn{3}{c}{Gaps} \\  
\cline{2-4} \cline{6-8} 
\colhead{} &\colhead{\#} & \colhead{Radius [au]} &  \colhead{Width [au]} &  \colhead{}  & \colhead{\#} & \colhead{Radius [au]}  & \colhead{Width [au]} 
}
\startdata
IM~Lup    & 28 & 146$_{89}^{302}$ & 93$_{76}^{134}$ && 18 & 194$_{124}^{275}$ & 42$_{16}^{56}$\\
GM~Aur    & 24 & 61$_{24}^{138}$ & 36$_{14}^{47}$ && 54 & 98$_{51}^{175}$ & 24$_{14}^{36}$\\
AS~209    & 28 & 63$_{39}^{106}$ & 56$_{42}^{72}$ && 15 & 70$_{42}^{87}$ &24$_{18}^{28}$  \\
HD~163296 & 44 & 108$_{40}^{233}$ & 54$_{40}^{100}$ && 29 & 97$_{87}^{203}$ & 26$_{19}^{34}$ \\
MWC~480   & 25 & 78$_{67}^{190}$ & 56$_{42}^{75}$ && 17 & 64$_{55}^{164}$ & 29$_{21}^{43}$ \\ \hline
Total     & 149 & 81$_{44}^{169}$ & 56$_{40}^{92}$ && 93 & 97$_{58}^{197}$ & 26$_{19}^{43}$\\
\enddata
\tablecomments{The distribution of radius and width for rings and gaps are given as the lower quartile, median, and upper quartile. We note that the widths are affected by beam convolution. See MAPS III for more details and a full listing of chemical substructures.}
\end{deluxetable*}

The diversity of chemical structures observed towards HD~163296 is seen towards the MAPS sample as a whole. Across the full sample, more than 200 emission rings, gaps, and plateaus are detected at all disk radii, from $\sim$10~au to 700~au. Table \ref{tab:ring_gap_stats} lists the numbers of rings and gaps identified towards each disk, considering only the 18 bright lines analyzed in MAPS III, as well as median and lower and upper quartiles for gap and ring radii and widths. Median ring and gap radii vary by a factor of 2--3 between different disks. Ring and gap widths range between $<$10~au (unresolved) and 200~au, but a majority of sub-structures are $<$100~au in width;  the median widths are 56 and 26~au for rings and gaps, respectively.  Finally, ring-gap contrasts vary substantially between lines, between different disk locations for the same line and disk, and between disks for the same line and disk radius.  The deepest gaps are close to empty with ring-gap contrasts of $>$90\%, but most gaps are more shallow, and the typical contrast is 10--30\%.

\begin{figure*}[ht!]
\centering
\includegraphics[scale=0.24]{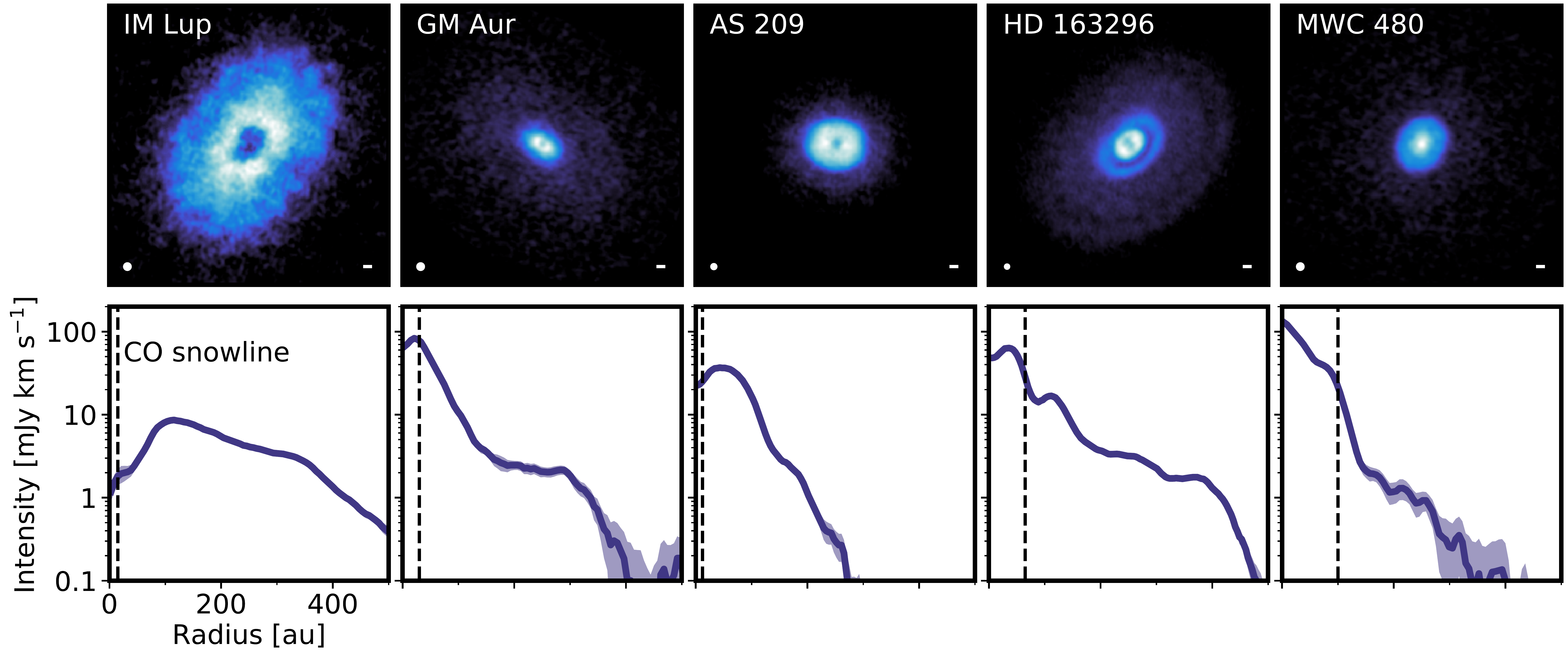}
\caption{Comparisons of emission morphologies (top row) and radial intensity profiles (bottom row) of the same molecular line, HCN J=3--2, F=3--2, between the five disks. The map sizes have been normalized to the same distance to facilitate comparison in morphology between sources. The dashed lines mark the estimated CO snowline locations towards each disk.  The images are shown with a powerlaw color stretch to emphasize the outer disk fainter structures, and normalized to the peak intensity in each panel. The synthesized beam is indicated in the bottom left corner, and a 20~au scalebar in the bottom right corners. \label{fig:comp-hcn}}
\end{figure*}

Figure \ref{fig:gallery} shows some examples of the gap and ring diversity observed across the disk sample.  This variation is further highlighted in Figure \ref{fig:comp-hcn} (lower panels), which shows the HCN 3--2 zeroth moment maps and radial profiles towards the MAPS disks. HCN emission ranges from faint and broad (IM Lup) to bright and compact (MWC 480), and from multi to single-ringed. It is currently unclear why the emission morphology is so diverse between the different disks, but we note that we purposefully picked disks with different dust structures, ages, and stellar properties, and it is therefore, perhaps, not a surprise that the chemical structures are different as well. 

Switching from the radial to the vertical dimension, MAPS IV presents the emission heights of CO 2--1 isotopologues, HCN and C$_2$H 3--2;  empirical 2D temperature models; and an inventory of vertical substructures. The extraction of emission surfaces was briefly described in \S3.4. The resulting emission height uncertainties strongly depend on the overall line strength as well as on disk radius; the uncertainties are high in the inner disk where we are resolution limited, and in the outer disk where we are sensitivity limited. At intermediate disk radii the uncertainties can be $<$10\% for the brightest lines, while it varies between 10 and 50\% for HCN towards HD~163296 (Fig. \ref{fig:imaging}). The CO isotopologue emission heights vary substantially between the five disks, which implies a diverse set of disk vertical density and temperature structures. The HCN and C$_2$H emission appears relatively flat, and towards the disks it could be constrained (AS~209, HD~163296, and MWC~480), the emission originates close ($z/r\sim0.1$) to the midplane. This suggests that the targeted HCN and C$_2$H lines enable us to probe organic abundances close to the planet-forming midplane. Identified vertical substructures are often associated with disk pressure minima, suggesting that they may be tracers of ongoing planet formation.

\subsection{Radial distributions of C, N, O, and S carriers, organics, ions, and deuterated molecules}

CO is the most abundant C and O-carrier observed by MAPS. Its column density radial profile and inferred depletion patterns towards the five MAPS disks are reported in \citet{Zhang21}, herein {\bf MAPS V}. CO is inferred to  be depleted in the warm molecular layers in all disks, but the degree of depletion varies substantially across and between disks; the nature and distributions of C and O reservoirs may be highly disk-dependent. The CO column density profiles display gaps and cavities. In MAPS V, we use the measured CO gap depths to infer gas gap depths, and further, to compare the relative depths of gas and dust gaps with expectations from planet-disk interaction models. The results are mixed, with some gas-to-dust gap ratios in good agreement with models, while others are off by almost an order of magnitude.

C$_2$H and HCN are potential probes of elemental ratios in disks; high C$_2$H abundances are only expected when the C/O ratio is elevated \citep{Bergin16,Miotello19}, and HCN formation rates depend on both the gas-phase C/O and N/O ratios \citep[e.g.,][]{Cleeves18}. Together with H$_2$CO, C$_2$H, and HCN are also probes of the disk organic feedstock chemistry.  \citet{Guzman21} or {\bf MAPS VI} present C$_2$H, HCN, and H$_2$CO column density profiles, and use the combined data to evaluate the organic reservoir, with focus on the inner 50 au of each disk.  Our multitransition data constrains C$_2$H to warmer disk layers and suggests that HCN is emitting from closer to the planet-forming midplane. 
There is substantial substructure in the column densities of all three molecules, including single and double rings, gaps and plateaus, some of which can be associated with proposed planet locations. 

In {\bf MAPS VII}, \citet{Bosman21_VII} use the CO and C$_2$H results from MAPS V and MAPS VI, respectively, to constrain the C/O ratios in three MAPS sources (AS 209, HD 163296, and MWC480) with dust and/or kinematic signs for forming planets. Surprisingly, they find that gas-phase C/O$\sim$2 across most of the disks, which implies that CO is not the dominant C-carrier in the warm molecular layer.  Alongside the constraints on the CO abundance, this suggests that 
putative Gas Giants assembling in these disks may be accreting envelopes with high C/O ratios and substellar O/H or metallicity, resulting in reducing atmospheres.%

One of the major dust gaps in the AS 209 disk coincides with substantial CO depletion. The origin of the CO depletion and its implications for planets forming in the gap is explored in more detail by \citet{Alarcon21} in {\bf MAPS VIII}. By combining thermochemical models and constraints on the disk pressure profile, MAPS VIII finds that the observed CO depletion is mainly caused by CO chemical processing rather than a local depletion of gas surface density. This limits the mass of the putative planet in the gap as the scale of the H$_2$ surface density perturbations must be less than that seen in the CO surface density profile.

Complementary to MAPS VI, which focuses on small organics, {\bf MAPS IX} \citep{Ilee21} presents the column density and abundance distributions of the larger organics CH$_3$CN, HC$_3$N, and, $c$-C$_3$H$_2$. Their column densities in the inner ($<$50 au) disk regions are $\sim$10\% compared to HCN or C$_2$H. Based on excitation temperatures of 20--40~K, CH$_3$CN observations probe organic reservoirs that reach down into the planet-forming midplane, and therefore provide constraints on the complex organic compositions of forming exocomets and exoplanets, including their prebiotically interesting nitrile budgets.

The origin of Solar System nitriles as well as the utility of deuterated molecules as chemical probes are the topics of {\bf MAPS X}, where \citet{Cataldi21} explore the D/H ratios of HCN and N$_2$H$^+$ in the MAPS disks. They find that the DCN/HCN ratio varies by orders of magnitudes across the disk sample, from $\sim$10$^{-1}$ in the outer regions of most disks, to 10$^{-3}$ in the inner 10s of au of the MWC 480 disk. By contrast, N$_2$D$^+$ is only detected outside of the CO snowline and the N$_2$D$^+$/N$_2$H$^+$ ratio is 10$^{-2}$--1. The results are best explained by the presence of two distinct deuterium fractionation channels (based on H$_2$D$^+$ and CH$_2$D$^+$) in disks, with implications towards the origin of deuterium enrichments within Solar System bodies, including in comets.

The final MAPS paper on nitriles, \citet{Bergner21} or {\bf MAPS XI}, uses CN column density profiles and CN/HCN column density ratios to explore disk photochemistry. The CN/HCN ratios increase with disk radius, reaching $\sim$100, such that CN is the dominant nitrile carrier at most disk radii. The CN/HCN ratio is also elevated in some gaps and appears to generally map out UV transparent disk regions. The high CN/HCN ratio observed across most of the disks suggests that photochemistry is key to understand the overall disk chemical composition. 

So far we have been discussing constraints on the C, N and O reservoirs, but a fourth important volatile element is sulfur (S), which in disks is most easily traced using CS. In {\bf MAPS XII}, \citet{LeGal21} use CS together with upper limits on SO to independently constrain the C/O ratio to $>$0.9 in the warm molecular layer ({\it cf.} MAPS VI and VII which instead make use of C$_2$H and CO observations). The H$_2$CS/CS ratio is measured in the MWC 480 disk and found to be high (on the order of unity), This ratio exceeds model predictions by two orders of magnitude, and indicates that sulfur participates in the disk organic chemistry at a higher level than expected. 

In the final paper in this section, \citet{Aikawa21} or {\bf MAPS XIII}, derive HCO$^+$ disk column density profiles and combine these data with N$_2$H$^+$ and N$_2$D$^+$ column densities from MAPS X to constrain the ionization fraction across the MAPS disks. The HCO$^+$ abundance, and hence the ionization level in the warm molecular layer, is relatively constant across most of the disks. HCO$^+$ is enhanced in some of the more pronounced disk gas gaps (as traced by CO), however, which suggests a higher ionization level in gaps compared to surrounding disk material.

\subsection{Disk dust and gas structures, masses, and kinematics}

While MAPS focuses on gas in disks, our multi-wavelength approach also resulted in new constraints on the dust population in the five MAPS disks. In {\bf MAPS XIV}, \citet{Sierra21} use multi-band continuum observations to constrain the dust size distributions, opacities, and surface densities. The grain sizes  often have local maxima in continuum rings, which is consistent with the prevailing theory of dust trapping due to pressure gradients. However, the inferred maximum grain sizes in the rings vary both between rings in the same disk, and between disks. Three of the disks are optically thick in the inner disk regions, and in these cases including scattering strongly affects the conclusions, including the inferred degree of grain growth in the inner disk. 

{\bf MAPS XV} \citep{Bosman21_XV} explores the gas and dust in the inner 20~au of disks at 3~au resolution using kinematic information from CO isotopologue line profiles. They use this technique to characterize  the dust gaps in the inner 20~au  of the AS~209 and HD 163296 disks, and find that  both  are gas-rich. The study also reveals that CO emission is substantially depleted in the inner 20~au of the IM Lup disk, which is best explained by high pebble fluxes from the outer to the inner disk. Such a high pebble flux should promote rapid formation of planetesimals through activation of the streaming instability, as well as efficient formation of planets through pebble accretion.

{\bf MAPS XVI} \citep{Booth21b} also constrains the properties of innermost disk regions, but this time by characterizing the large-scale outflow identified towards HD~163296 \citep{Klaassen13} using multi-wavelength CO isotopologue observations. The emission is interpreted  as an MHD driven disk wind, which constrains the launch radius to 4~au. The angular momentum removed by this wind is sufficient to drive the current disk accretion rate, which removes the need to invoke turbulent viscosity in this region of the disk. This has profound implications for the physics and evolution of the inner ($<$10~au) planet-forming zone of the disk.

The HD~163296 disk is also studied by \citet{Calahan21} in {\bf MAPS XVII}. With a detailed model fit to all MAPS observations of CO isotopologues, MAPS XVII reveals the global thermal structure of the disk and constrains the planet-forming mass reservoir. The study also shows that the gaps associated with potential planets should be  modestly warmer than their surroundings. This implies that we should  expect subtle changes in line brightness temperatures in gas-depleted gaps carved out by massive planets as compared to gas-rich gaps.

{\bf MAPS XVIII} explores the kinematic substructure of the disks around Herbig Ae stars HD~163296 and MWC~480, and uses the identified velocity patterns to infer the probable presence of nascent planets \citep{Teague21}. In particular, MWC~480 presents a tightly wound spiral pattern and localized velocity perturbations, which {\bf can be explained} by a gas giant at $\sim$250~au, i.e.,  well beyond the typical locations of gas-giant planet formation in models. The HD~163296 analysis further strengthens previous kinematic claims of multiple gas giants forming in this massive disk. As a whole, this study further develops  the potential of millimeter gas observations as a tool to find and estimate masses of the youngest planets.

 \citet[][from now on {\bf MAPS XIX}]{Huang21} presents the curious  gas-structures associated with the GM Aur disk. In addition to a large Keplerian disk, CO emission towards GM Aur presents spiral arms in the outer disk, a tail extending southwest, and diffuse structures surrounding the north side of the disk. One explanation for these large-scale, complex CO structures is late infall of a remnant envelope or cloud material, which may change the overall disk mass budget for planet formation. 
 Further, ongoing interactions of disks with their natal clouds imply that planet-forming material may be of a range of chemical ages, and therefore more chemically diverse than typically assumed.

{\bf MAPS XX} presents a detailed model of the same GM Aur disk to constrain its gas mass and surface density \citep{Schwarz21}. The analysis includes all MAPS CO data as well as archival data, for a complete data set of 11 ALMA CO line image cubes, and HD 1--0 from {\it Herschel}. The best fit is obtained if the disk is cold and massive, $\sim$0.2 M$_\odot$. This would make the GM Aur disk one of the most massive disks discovered to date. The derived disk gas surface density suggests that the disk may be gravitationally unstable in one of the dust rings, between 70 and 120~au.

\section{Discussion}\label{sec:disc}

MAPS observations cover disk scales from $<1$0 to $\sim$1000 au. All of these scales are relevant to planet formation. Winds extending 1000s of au from the HD~163296 disk probe the physical conditions at their launch point in the innermost disk regions (MAPS XVI). Spirals extending 100s of au beyond the pebble disk likely trace infalling material onto the GM Aur disk which may affect the mass, chemical make-up, and stability of the pebble disk where planet formation is likely ongoing (MAPS XIX and XX). At intermediate disk radii of 10--200 au, resolved dust and molecular substructure allow us to assess the relationship between dust (MAPS XIV), gas (MAPS V, XV, and XVII) and chemical substructure (MAPS III -- VIII), and how they all relate to ongoing planet formation (MAPS XVIII). These intermediate disk radii, which are well resolved in MAPS, are also ideal to explore how disk UV fields, temperatures, ionization, and elemental ratios affect chemistry that may later become incorporated into planet-building solids (MAPS IX -- XIII). We note that the disk region most directly relevant to planet formation according to planet formation models is the inner 10s of au of the disk midplane \cite[e.g.,][]{Pollack96,Raymond14,Johansen17}. This coincides with the disk regions interior to the CO snowline, which is estimated to occur at 12--100 au in our disks (MAPS V). Interior to the CO snowline, we can in theory access the gas composition all the way down to the disk midplane, resulting in direct constraints on the chemistry of planet-forming material.

A detailed discussion of the respective results and their impact on our understanding of planet formation can be found in the individual MAPS papers. Here we provide some discussion relevant for the project as a whole on the chemistry of the planet-forming midplane and the disk organic reservoirs in the main planet-forming disk regions, on interactions between planet formation and chemistry, and on chemical probes of disk properties and nascent planets.

\subsection{Chemistry in the planet-forming midplane}

Planets form in the midplanes of disks, and observations that directly probe midplane material are therefore especially relevant for predicting planet and planetesimal compositions \citep[though see][for evidence that planets can also accrete material from elevated disk layers]{Morbidelli14,Teague19}. An important question for a program on the chemistry of planet formation is then: How well can high-resolution millimeter observations probe the disk midplane? 

Interior to the CO snowline, CO and related molecules are present in the gas-phase at all disk heights and observations of their respective molecular lines may in principle be used to assess the volatile molecular reservoirs and the local disk conditions down to the disk midplane. MAPS IV shows that C$^{18}$O generally originates at $z/r<$0.1, close to the midplane in the inner 100~au of disks. In the three disks where the vertical distribution of HCN and C$_2$H could be constrained, they too emit from $z/r\lesssim0.1$ interior to the snowline. The vertical emission layers of molecules are also constrained by the molecular excitation temperature profiles, which can be used to map out their emitting regions if the overall temperature structure is known (see MAPS IX). Interior to the CO snowline, derived excitation temperatures place most molecules (HCN, DCN, C$_2$H, CS, and CH$_3$CN) close to the midplane (MAPS VI, IX, X, and XII). The column densities reported in MAPS V-XIII interior to the CO snowline are then largely probing the chemical conditions in, or in close vicinity to the planet-forming midplane.

Exterior to the CO snowline, there are few molecules left in the gas-phase in the disk midplane.  A handful of molecules, such as N$_2$H$^+$, N$_2$D$^+$ (probed by MAPS X), and H$_2$D$^+$, should still be in the gas-phase, and their emission could be used to probe the outermost planet-forming disk regions, including the ongoing deuterium fractionation chemistry \citep[][MAPS X]{Willacy07,Huang15,Salinas17,Aikawa18}, and ionization levels (MAPS XIII). The midplane elemental and organic compositions between the CO snowline and the pebble disk edge are, however, mostly hidden from view. In the outermost disk regions, beyond the pebble disk edge, UV radiation may penetrate all the way to the disk midplane, releasing some of the frozen out molecules back into the gas-phase \citep{Oberg15b,Cleeves15b}, providing renewed access to the disk chemical composition. This scenario is confirmed by MAPS X and XI. The relevance of these outer disk regions to planet formation is unclear, but the discovery of potential infall onto the outer disk from the surrounding environment in MAPS XIX suggests that the outer disk may not infrequently act as a chemical bridge between the natal cloud environment and the planet-forming regions of the disk. 

\subsection{Organic reservoirs interior to 50~au}
 
 \begin{deluxetable*}{lccccc}
\tablecaption{Minimum volatile organic reservoirs $<$50~au$^\dagger$ }
\label{tab:organics}
\tablehead{
Source&\colhead{Water}   &\colhead{Nitriles}   &\colhead{All organics}  &\colhead{Nitriles}   &\colhead{All organics}\\
&\colhead{[\# Earth Oceans]} &\colhead{[\# Earth Oceans]} & \colhead{[\# Earth Oceans]}& \colhead{[\% H$_2$O]} & \colhead{[\% H$_2$O]} 
}
\startdata
IM Lup      &240,000    &0.20   &2.0    &$8.1\times10^{-5}$&$8.3\times10^{-4}$\\
GM Aur      &25,000 &150    &170    &0.60   &0.69\\
AS 209      &1,300  &21 &47 &1.6    &3.5\\
HD 163296   &49,000 &340    &390    &0.70   &0.79\\
MWC 480     &96,000 &440    &460    &0.46   &0.48\\
\hline
Comet 67P  &   &   &   &0.17$^\star$   &0.83$^\diamond$\\
\enddata
\tablenotetext{}{$^\dagger$ Assuming a 1000-to-1 ice-to-gas ratio of volatile organics.}
\tablenotetext{}{$^\star$ Summing up the abundances of all organics with a CN group in \citet{Drozdovskaya19}.}
\tablenotetext{}{$^\diamond$ Summing up the abundances of all volatile organics listed in \citet{Drozdovskaya19}.}
\end{deluxetable*}

Within MAPS, we observe a number of small and larger organic molecules: HCN, C$_2$H, H$_2$CO, $c$-C$_3$H$_2$, CH$_3$CN, and HC$_3$N (MAPS VI and IX). The total amount of these organics in the gas-phase in the inner 50~au of the disks is substantial, up to half an Earth Ocean (MAPS VI and IX). However, the true volatile organic reservoir should be considerably larger. First, ALMA can only detect a subset of the organics observed in the remnants of our own disk, i.e., in comets; 48 different organic molecules have been found towards comet 67P to date \citep{Altwegg19},  while only 12 (10 at millimeter wavelengths) have been detected in protoplanetary disks \citep{McGuire18}. Second, and more importantly, the gas organic reservoir directly probed by observations is likely small compared to the organic ice reservoir. Most of the organics are constrained to have excitation temperatures of 30--40~K. This is below the expected freeze-out temperatures of the same molecules, with the possible exception of C$_2$H \citep{Wakelam12,Wakelam17,Bertin17}, which suggests that the main organic reservoir will be icy grains and pebbles. 

Calculating the precise balance between gas and ice abundances at any one disk location would require a dedicated modeling effort that takes into account both thermal and non-thermal desorption mechanisms. In the meantime, we can use previous models to estimate the order of magnitude ice-to-gas ratio. \citet{Walsh14} report gas and grain surface column densities for their fiducial disk as a function of radius, and find H$_2$CO and CH$_3$CN  ice-to-gas ratios of $\sim$10$^3$, and a HC$_3$N ratio of 4 at 20~au (taken here to be representative for the disk interior to 50~au). \citet{Ruaud19} find higher ratios at 20--50~au of $\sim$10$^5$ for HCN, $\sim$10$^4$ for H$_2$CO,  $\sim$10$^7$ for CH$_3$CN, and $\sim$10$^5$ for HC$_3$N. \citet{Oberg15} found HCN ice-to-gas ratios of 10$^2$--10$^6$ and CH$_3$CN ice-to-gas ratios of 10$^4$--10$^8$, dependent on  the disk turbulence level. If we conservatively adopt a ice-to-gas ratio of 10$^3$ in the disk, we estimate a minimum volatile organic reservoir of 2--460 Earth oceans interior to 50~au in the MAPS disks (Table \ref{tab:organics}), including only the molecules that are covered by MAPS. The total reservoir of the prebiotically-relevant nitriles \citep{Patel15} in the same disk region varies between 0.2 and 440 Earth Oceans. The basic building blocks of a pre-biotic origins of life chemistry are plentiful in the comet- and planet-forming regions of protoplanetary disks.

We can compare the abundances of organics to the young Solar System by normalizing them to water. We estimate the total water mass $<$50~au from the models of each disk in MAPS V, and list these estimates in Table \ref{tab:organics}. These should be seen as order of magnitude estimates, since they depend sensitively on the  inferred surface densities in the inner regions of these disks, as well as on the fraction of the inherited water that was chemically converted into other species during the disks' lifetime -- in our models all disks start out with water abundances of $1.8\times10^{-4}$ n$_{\rm H}$, but after 1 Myr, the abundances vary between $0.25-1.8\times10^{-4}$ n$_{\rm H}$. In 4/5 disks, both the nitrile and total organics abundances are $\sim$1\% compared to H$_2$O. For nitriles, this is in good agreement with Solar System comet values \citep{Mumma11,Altwegg19,Drozdovskaya19}. We note that the nitrile/(total organics) ratio in our disks is considerably higher than in most comets (67P excluded), which suggests that while the {\sl in situ} production of nitriles in disks may account for much of the final nitrile budget, other sources are needed for oxygen-bearing organics. They may be in large part inherited from the previous star formation stages \citep{Drozdovskaya19,Oberg21_Review}.

\subsection{Links between dust and chemical substructures}

One of the fundamental questions targeted by MAPS is the link between ongoing planet-formation, as traced by millimeter continuum substructures, and gas and chemical substructures. More specifically, with MAPS we begin to address the following questions: How is the gas distributed across disks with dust substructures?  Is the chemistry in gaps distinct compared to the surrounding disks? Are line emitting temperatures affected by dust gaps? Is there a link between snowlines and other chemical substructures?

Despite numerous pieces of theoretical and observational evidence for CO redistribution and depletion in disks \citep{Reboussin15,Miotello17, Yu17,  Bosman18,Schwarz18,Dodson-Robinson18,Zhang19,Krijt20}, CO remains the overall best tracer of the gas distribution in disks \citep{Molyarova17}. We find that the there is some correlation between dust gaps and CO column density gaps, but there are also several deep dust gaps with no visible CO and, by inference, gas depletion in them (MAPS V). This suggests that not all dust gaps are created by the same mechanism, and that the inferred gas depth profiles could be used to constrain the origin of the dust gap, including the mass of any putative planet \citep{Kanegawa15,Zhang18}. The use of CO to trace the gas properties of dust gaps is complicated, however, by evidence, of CO depletion. In particular, the large C$_2$H abundances observed in several disks are best explained by a substantial CO removal from the gas. This implies that the observed CO column density decreases in some dust gaps may in part be due to  chemical CO depletion rather than gas depletion (MAPS VII and VIII), which entails that the inferred gas depletions in dust gaps are really upper limits. 

Based on theoretical models, the emitting temperature of gas in gaps may either be cold compared to nearby disk regions because of gas-dust decoupling, shadowing or an emitting layer that is closer to the disk midplane  \citep{Jang-Condell08,Facchini18,Alarcon20a}, or warm  because the gap heating-cooling balance is different compared to the surrounding disk \citep[][MAPS XVII]{vanderMarel17,Teague19c,Alarcon20a}. In MAPS, most gaps are not associated with a change in temperature as  measured by line intensities of optically thick lines (MAPS IV), but there are some exceptions. One of the gaps in the HD~163296 disk and one of the gaps in the MWC 480 disk are each characterized by a lower CO excitation temperatures compared to the surrounding disk (MAPS IV, XVIII). Because of the relatively low resolution of Band 3 observations, we were unable to constrain changes in the excitation temperature across gaps using multi-line analysis.

There is also no consistent association between dust gaps and rings, and the gaps and rings seen in molecular lines other than CO.  At least this is the case if comparing dust and gas substructures regardless of the contrast or width of the dust gap or ring. If we instead focus on the innermost dust gap of each disk that is clearly seen in the MAPS continuum data (D116 in IM Lup, D15 in GM Aur, D61 in AS 209, D49 in HD~163296, and D76 in MWC 480 \citep[see][for notation]{Huang18b}) a somewhat more coherent picture emerges, where dust depletion is associated with an excess of nitriles and hydrocarbons (MAPS III), indicative of oxygen depleted gaps (MAPS VII). It is curious, however, that in, e.g.,\ HD~163296 the very next major dust gap is instead associated with a gap in nitriles and hydrocarbons, indicating that gaps that appear similar in continuum may be diverse in their gas and chemistry. The cause of this diversity is unclear and future dedicated modelling efforts are needed to better understand the link between dust sculpting and  chemistry.

Snowlines may cause chemical substructures because they dramatically change the molecular abundance of the molecule in question, and also because they affect the gas elemental composition. The MAPS resolution enables us to probe chemical changes across the CO and N$_2$ snowlines. However, we only have independent constraints on CO snowline locations in the GM Aur and HD~163296 disks, and on the N$_2$ snowline towards the GM Aur disk \citep{Qi15,Qi19}. With these caveats in mind, it is still noteworthy that we see no consistent link between inferred CO snowline locations and dust or chemical substructure (MAPS III).  This is consistent with previous results of \citet{Huang18b} and \citet{Long18} comparing dust substructure and inferred snowline locations towards larger samples of disks, as well as a recent in-depth chemical exploration of the transition disk around AB Aur \citep{Riviere20}. These results together suggest that either our current estimates of snowline locations are inaccurate\footnote{Temperature profiles of disks depend sensitively on assumptions about small grain population, which are not well constrained, and this together with a range of possible CO freeze-out temperatures make snowline locations difficult to infer.}, or that the impact of snowlines on the radial chemical profiles is modest. The latter may point to a chemical disconnect between the midplane and warm molecular layers beyond the CO snowline. This is supported by the observation that many chemical substructures in the outer disk also do not coincide with dust substructures (MAPS III), which also are located in the midplane.

Since many chemical substructures cannot be explained by dust substructures or snowlines, we need to also consider other disk characteristics that can cause chemical substructures. For example, gradients in disk properties such as UV radiation flux (MAPS XI), ionization, and temperature may result in radial locations that are particularly favorable or unfavorable to the formation of a specific molecule, and chemical substructures may emerge as a powerful probe of such disk characteristics.

\subsection{Molecular probes of planet formation}

In addition to constituting probes of disk chemistry, molecular line emission can be used to constrain  several disk properties that are important to planet formation, such as gas mass, kinematics, temperature, snowline locations, and ionization, and have been proposed to enable detections of nascent planets. Deploying disk kinematic planet detection methods \citep{DiskDynamics20}, the MAPS data have resulted in the possible identification of a new planet in MWC 480, and additional dynamical data on the previously proposed planets in the HD~163296 disk (MAPS XVIII). On larger scales, MAPS XVI characterized the dynamics of the HD~163296 molecular wind, which also constrains the dynamics of the innermost regions of the disk. Towards GM Aur, MAPS XIX found evidence of an ongoing interaction between the disk and surrounding material with implications for the disk composition, structure, and dynamics. These interactions with the environment resemble what is seen towards younger sources \citep[see][for recent examples]{Pineda20,Alves20}. We note that all studies of dynamics in MAPS (XVI, XVIII, XIX) are based on $^{12}$CO and $^{13}$CO lines. Whether it will be possible to identify forming planets and other dynamical phenomena using lines of less abundant molecules \citep[e.g.,][]{Cleeves15b} remains to be seen, but it may very well require deeper observations than was pursued with MAPS, or observations of other strong lines such as CN, CS and HCO$^+$ in Band 6 or Band 7.

The disk gas mass places fundamental constraints on what kind of planetary system can form in a disk, and statistics on gas masses are needed to provide an interpretive framework for exoplanet demographics. There is currently no single reliable tracer of disk gas mass, i.e.,\ commonly used ones such as CO isotopologues \citep{Williams14,Miotello16,Bergin17}, HD \citep{Bergin13,McClure16}, and dust edge locations \citep{Powell19} all have their limitations. MAPS XX uses two of the above constraints together with a grid of thermochemical models to determine the GM Aur gas mass. With no far-IR mission currently planned, 
 and considering that the number of disks for which {\it Herschel} HD constraints exist is small, we will need to develop additional disk mass tracers to obtain statistics on disk masses.

The disk temperature structure determines the locations of snowlines, and affects formation and migration of planets. Despite a long recognition of its importance, it has been elusive to constrain. Spatially resolved optically thick CO lines have been used in a handful of case studies to constrain the disk temperature structure \citep{Dartois03,Zhang17,Isella18,Pinte18}. In MAPS, we developed a semi-automatic workflow to map out 2D disk temperatures (MAPS IV). This works well and is an attractive approach to obtain model independent disk temperature structures. We note that the utility of this method depends strongly on the inclination of the disk and the spatial resolution, which will limit how quickly we can build up a large library of such empirical disk temperature. Once such a library exists, disk thermochemical models are key to provide the interpretative framework as is illustrated by MAPS XVII. 

Snowlines can sometimes be probed using the volatile of interest \citep[e.g.,][]{Schwarz16}, but most of the time we rely on chemical tracers such as N$_2$H$^+$ or N$_2$D$^+$ for CO and N$_2$ snowlines \citep{Qi13c,Qi19,Cataldi21}, and HCO$^+$ for H$_2$O snowlines \citep{Jorgensen13,Visser15,vantHoff18a,vantHoff18b,Leemker21}. These chemical tracers are  needed to resolve ambiguities between snowlines and changes in gas surface densities, and because many common disk volatiles are difficult or impossible to observe at millimeter wavelengths, including H$_2$O, CO$_2$, NH$_3$ and N$_2$. 

Ionization levels in disks are most directly probed by observing the lines from major ions. MAPS included lines from two ions, HCO$^+$ and N$_2$D$^+$, that are proposed to be major charge carriers in the warm molecular and cold midplane layers, respectively. We note that based on MAPS, N$_2$D$^+$ lines appear generally detectable in large disks with a few hours of integration, providing a window into ionization in the outer disk midplane. The 1--0 line of HCO$^+$ was readily detected in all disks, while the $^{13}$C isotopologue was more challenging to observe, but still detected in some disks (MAPS XIII). Together the two isotopologues  provided good constraints on ionization in the outer disk warm molecular layer. What remains to be constrained is the ionization level in the inner 20--40~au in disks, which will likely require additional observations of the brighter, higher frequency HCO$^+$ lines.

\section{Conclusions}
\label{sec:conc}

MAPS set out to survey the distributions of common molecules in five disks around three T Tauri and two Herbig Ae stars  -- IM Lup, GM Aur, AS 209, HD~163296 and MWC~480 -- at unprecedented detail with ALMA, and to develop an interpretative framework for the observations. Our main findings are

\begin{enumerate}
    \item The imaging process and the production of higher level data products required substantial development compared to previous best-practices, and we recommend that other projects pursuing high-resolution spectral-imaging adopt and build on our integrated workflow, which is described in detail in MAPS II. 
    \item All molecules surveyed by MAPS at high spatial resolution (7--30 au) present some substructures, resulting in over 200 identified rings, gaps and plateaus. This suggests that disk gas properties, including the chemical composition, vary substantially on small scales in disks, and therefore that planets may form in chemically distinct environments.
    \item The MAPS spatial scales also enabled constraints on the vertical emission and temperature profiles. The resulting empirical 2D temperature structures are key to anchoring disk models, and deriving the temperatures of the planet forming disk-midplanes.
    \item The CO gas-phase abundances vary dramatically across the MAPS disks, and are depleted by 1--2 orders of magnitude at most disk locations compared to the interstellar canonical value. The C/H  and O/H ratios are therefore substellar. In addition the C/O gas-phase ratio is elevated above unity in much of the disks, with implications for predicted planet envelope compositions. The MAPS molecular inventory also includes probes of the N and S reservoirs, and we present some initial constraints, but more theoretical work is needed to derive quantitative abundance patterns. 
    \item MAPS covered six small and mid-sized organic molecules: HCN, C$_2$H, H$_2$CO, HC$_3$N, CH$_3$CN, and $c$-C$_3$H$_2$. Their radial distributions are different, indicating a changing organic inventory with disk radius. Within 50~au, the inferred organic inventory is large, corresponding to many Earth oceans, and planets forming in these disks are assembling in an organically rich environment.
    \item MAPS included probes of deuterium fractionation, photochemistry, and ionization, which are all important to interpret the volatile reservoirs in the Solar System, and predict the volatile content of exoplanets. In particular, MAPS has enabled high-resolution constraints on the distribution of deuterium chemistry, photochemical products, and ionization levels, including their distributions at radii directly relevant for planet and comet formation. 
    \item While not an original motivation for MAPS, the MAPS CO data have revealed multiple pieces of evidence for the dynamic nature of these protoplanetary disk systems including disk winds, gas spirals, and azimuthal asymmetries that we could connect with ongoing planet formation. Furthermore, MAPS has enabled us to further develop kinematic planet detection methods, and resulted in the possible localization of a planet in the MWC 480 disk.
\end{enumerate}

Together these results demonstrates the utility of deep, high-resolution ALMA observations of molecular lines in disks to explore the chemistry that affects and probes planet formation. Going forward, we will need to expand this approach to larger samples of representative disks around a range of stars, including the frequently planet-hosting M Dwarfs, to obtain statistically meaningful constraints on the chemistry of planet formation. 

\acknowledgments

 The authors thank the anonymous referee for valuable comments that improved both the content and presentation of this work. This paper makes use of the following ALMA data: ADS/JAO.ALMA\#2018.1.01055.L. ALMA is a partnership of ESO (representing its member states), NSF (USA) and NINS (Japan), together with NRC (Canada), MOST and ASIAA (Taiwan), and KASI (Republic of Korea), in cooperation with the Republic of Chile. The Joint ALMA Observatory is operated by ESO, AUI/NRAO and NAOJ. The National Radio Astronomy Observatory is a facility of the National Science Foundation operated under cooperative agreement by Associated Universities, Inc.

K.I.\"O. acknowledges support from the Simons Foundation (SCOL \#321183) and an NSF AAG Grant (\#1907653). V.V.G. acknowledges support from FONDECYT Iniciaci\'on 11180904 and ANID project Basal AFB-170002.  C.W. acknowledges financial support from the University of Leeds, STFC and UKRI (grant numbers ST/R000549/1, ST/T000287/1, MR/T040726/1).  Y.A. acknowledges support by NAOJ ALMA Scientific Research Grant Code 2018-10B and 2019-13B, and Grant-in-Aid for Scientific Research 18H05222 and 20H05847. E.A.B. acknowledges support from NSF AAG Grant \#1907653. C.J.L. acknowledges funding from the National Science Foundation Graduate Research Fellowship under Grant DGE1745303. S.M.A. and J.H. acknowledge funding support from the National Aeronautics and Space Administration under Grant No. 17-XRP17 2-0012 issued through the Exoplanets Research Program. J.B. acknowledges support by NASA through the NASA Hubble Fellowship grant \#HST-HF2-51427.001-A awarded  by  the  Space  Telescope  Science  Institute,  which  is  operated  by  the  Association  of  Universities  for  Research  in  Astronomy, Incorporated, under NASA contract NAS5-26555. J.B.B. acknowledges support from NASA through the NASA Hubble Fellowship grant \#HST-HF2-51429.001-A, awarded by the Space Telescope Science Institute, which is operated by the Association of Universities for Research in Astronomy, Inc., for NASA, under contract NAS5-26555.  Y.B. acknowledges funding from ANR (Agence Nationale de la Recherche) of France under contract number ANR-16-CE31-0013 (Planet-Forming-Disks).
A.S.B acknowledges the studentship funded by the Science and Technology Facilities Council of the United Kingdom (STFC).
J.K.C. acknowledges support from the National Science Foundation Graduate Research Fellowship under Grant No. DGE 1256260 and the National Aeronautics and Space Administration FINESST grant, under Grant no. 80NSSC19K1534. G.C. is supported by NAOJ ALMA Scientific Research
grant No. 2019-13B.  L.I.C. gratefully acknowledges support from the David and Lucille Packard Foundation and Johnson \& Johnson's WiSTEM2D Program.  IC was supported by NASA through the NASA Hubble Fellowship grant HST-HF2-51405.001-A awarded by the Space Telescope Science Institute, which is operated by the Association of Universities for Research in Astronomy, Inc., for NASA, under contract NAS5-26555.  J. H. acknowledges support for this work provided by NASA through the NASA Hubble Fellowship grant \#HST-HF2-51460.001-A awarded by the Space Telescope Science Institute, which is operated by the Association of Universities for Research in Astronomy, Inc., for NASA, under contract NAS5-26555. J.D.I. acknowledges support from the Science and Technology Facilities Council of the United Kingdom (STFC) under ST/T000287/1.  N.T.K. acknowledges support provided by the Alexander von Humboldt Foundation in the framework of the Sofja Kovalevskaja Award endowed by the Federal Ministry of Education and Research.  R.L.G. acknowledges support from a CNES fellowship grant.  Y.L. acknowledges the financial support by the Natural Science Foundation of China (Grant No. 11973090).  F.L. acknowledges support from the Smithsonian Institution as a Submillimeter Array (SMA) Fellow. F.M. acknowledges support from ANR of France under contract ANR-16-CE31-0013 (Planet Forming Disks)  and ANR-15-IDEX-02 (through CDP ``Origins of Life"). H.N. acknowledges support by NAOJ ALMA Scientific Research Grant Numbers 2018-10B and Grant-in-Aid for Scientific Research 18H05441. L.M.P.\ acknowledges support from ANID project Basal AFB-170002 and from ANID FONDECYT Iniciaci\'on project \#11181068. K.R.S. acknowledges the support of NASA through Hubble Fellowship Program grant HST-HF2-51419.001, awarded by the Space Telescope Science Institute,which is operated by the Association of Universities for Research in Astronomy, Inc., for NASA, under contract NAS5-26555. A.S. acknowledges support from ANID/CONICYT Programa de Astronom\'ia Fondo ALMA-CONICYT 2018 31180052. R.T. acknowledges support from the Smithsonian Institution as a Submillimeter Array (SMA) Fellow. T.T. is supported by JSPS KAKENHI Grant Numbers JP17K14244 and JP20K04017. Y.Y. is supported by IGPEES, WINGS Program, the University of Tokyo. M.L.R.H. acknowledges support from the Michigan Society of Fellows.  A.R.W. acknowledges support from the Virginia Space Grant Consortium and the National Science Foundation Graduate Research Fellowship Program under Grant No. DGE1842490.  K.Z. acknowledges the support of the Office of the Vice Chancellor for Research and Graduate Education at the University of Wisconsin – Madison with funding from the Wisconsin Alumni Research Foundation, and support of the support of NASA through Hubble Fellowship grant HST-HF2-51401.001. awarded by the Space Telescope Science Institute, which is operated by the Association of Universities for Research in Astronomy, Inc., for NASA, under contract NAS5-26555.  

\vspace{5mm}
\facilities{ALMA}


\software{Astropy \citep{Astropy18}, \texttt{bettermoments} \citep{Teague18_bettermoments}, CASA \citep{McMullin07}, \texttt{disksurf} (\url{https://github.com/richteague/disksurf}), \texttt{GoFish} \citep{Teague19b}, Matplotlib \citep{Hunter07}, NumPy \citep{vanderWalt_etal_2011}}

\appendix

\section{Observational Details}
\label{app:obs}

Tables \ref{tab:obs-b3} and \ref{tab:obs-b6}  list all ALMA  executions carried out as a part of MAPS. Table \ref{tab:obs-b3} lists the Band 3 executions, and Table \ref{tab:obs-b6} summarizes the Band 6 executions. These Tables include observing dates, number of antennas, on source integration times, baselines, observatory-estimated spatial resolution, maximum recoverable scale, and phase and flux calibrators. We note that GM Aur and MWC 480 were always observed in the same execution with the integration time split evenily between the two sources. The total number of executions was 90. Each execution included 36--52 minutes of on source integration (with the exception of short 11-17 min integrations).\\

\begin{deluxetable*}{llcccccccc}
\tablecaption{Details of Band 3 observations\label{tab:obs-b3}}
\tablehead{
\colhead{Setting}    &\colhead{Target}  &\colhead{Date}   &\colhead{\# Ant.} & \colhead{Int.}    & \colhead{Baselines}	 & \colhead{Res.}& \colhead{Max Scale}& \colhead{Phase Cal.}& \colhead{Flux Cal.}\\
 \colhead{} & \colhead{}    &\colhead{}    & \colhead{} &
\colhead{[min]}    &\colhead{[m]} &\colhead{}&\colhead{}& \colhead{}& \colhead{} 
}
\startdata
B3-1  &IM Lup  &2018-10-29  &48 &37 &15--1398   &0\farcs6  &8\farcs5   &J1610-3958 &J1427-4206\\
&&2019-08-20  &43 &38 &41--3396   &0\farcs22  &3\farcs7   &J1610-3958 &J1517-2422\\
&&2019-08-21  &44 &38 &41--3638   &0\farcs22  &3\farcs6   &J1610-3958 &J1517-2422\\
\hline
&GM Aur/MWC 480  &2018-12-13  &42 &46 &15--783   &1\farcs0  &13\arcsec   &J0438+3004 &J0510+1800\\
&&2018-12-15  &42 &47 &15--740   &1\farcs0  &12\arcsec   &J0438+3004 &J0510+1800\\
&&2019-08-31  &43 &46 &38--3396   &0\farcs23  &3\farcs8   &J0459+3106 &J0510+1800\\
&&2019-09-02  &43 &52 &38--3638   &0\farcs23  &3\farcs7   &J0438+3004 &J0510+1800\\
&&2019-09-02  &31 &52 &38--3638   &0\farcs23  &3\farcs8   &J0512+2927 &J0510+1800\\
&&2019-09-04  &43 &50 &38--3638   &0\farcs23  &3\farcs8   &J0438+3004 &J0510+1800\\
&&2019-09-04  &43 &50 &38--3638   &0\farcs23  &3\farcs8   &J0438+3004 &J0510+1800\\
\hline
&AS 209  &2018-10-26  &49 &36 &15--1398   &0\farcs6  &8\farcs6   &J1733-1304 &J1517-2422\\
&&2019-08-23  &43 &38 &43--3396   &0\farcs23  &3\farcs3   &J1653-1551 &J1550+0527\\
&&2019-08-24  &44 &38 &43--3396   &0\farcs22  &3\farcs2   &J1653-1551 &J1517-2422\\
&&2019-09-04  &46 &38 &38--3144   &0\farcs23  &4\farcs0   &J1653-1551 &J1517-2422\\
&&2019-09-04  &48 &38 &38--3637   &0\farcs23  &3\farcs8   &J1653-1551 &J1517-2422\\
\hline
&HD 163296  &2018-10-22  &48 &36 &15--1398   &0\farcs6  &8\farcs4   &J1743-1658 &J1924-2914\\
& &2019-08-23  &43 &37 &41--3144   &0\farcs23  &3\farcs7   &J1755-2232 &J1924-2914\\
& &2019-08-24  &45 &37 &41--3396   &0\farcs23  &3\farcs3   &J1755-2232 &J1517-2422\\
& &2019-08-25  &45 &37 &41--3396   &0\farcs23  &3\farcs8   &J1755-2232 &J1924-2914\\
& &2019-09-04  &46 &37 &38--3144   &0\farcs23  &4\farcs0   &J1755-2232 &J1924-2914\\
& &2019-09-05  &49 &37 &38--3638   &0\farcs22  &3\farcs8   &J1755-2232 &J1924-2914\\
\hline\hline
B3-2  &IM Lup   &2018-11-06 &47 &37 &15--1398   &0\farcs60  &9\farcs7   &J1610-3958 &J1427-4206\\
&&2019-08-22 &45 &38 &41--3638  &0\farcs23  &3\farcs6   &J1610-3958 &	J1517-2422\\
&&2019-08-22 &46 &38 &41--3638  &0\farcs23  &3\farcs8   &J1610-3958 &	J1517-2422\\
\hline
&GM Aur/MWC 480  &2018-12-15  &45 &34 &15--740   &1\farcs0  &12\arcsec   &J0438+3004 &J0510+1800\\
&&2018-12-15  &45 &34 &15--740   &1\farcs0  &12\arcsec   &J0438+3004 &J0510+1800\\
&&2019-04-15  &44 &34 &15--783   &1\farcs0  &14\arcsec   &J0438+3004 &J0510+1800\\
&&2019-08-30  &48 &49 &38--3396   &0\farcs24  &3\farcs9   &J0438+3004 &J0237+2848\\
&&2019-08-30  &48 &49 &38--3396   &0\farcs24  &3\farcs9   &J0438+3004 &J0510+1800\\
&&2019-08-30  &47 &49 &38--3396   &0\farcs23  &3\farcs9   &J0438+3004 &J0510+1800\\
&&2019-09-01  &47 &11 &38--3338   &0\farcs23  &3\farcs8   &J0519+2744 &J0510+1800\\
&&2019-09-01  &52 &49 &38--3638   &0\farcs23  &3\farcs8   &J0438+3004 &J0510+1800\\
\hline
&AS 209  &2018-10-27  &48 &36 &15--1398   &0\farcs6  &8\farcs7   &J1733-1304 &J1517-2422\\
&&2019-08-23  &44 &37 &41--3638   &0\farcs23  &3\farcs9   &J1653-1551 &J1517-2422\\
&&2019-08-23  &44 &37 &41--3396   &0\farcs23  &3\farcs6   &J1653-1551 &	J1924-2914\\
&&2019-09-17  &41 &37 &41--3638   &0\farcs24  &3\farcs9   &J1653-1551 &	J1924-2914\\
\hline
&HD 163296  &2018-10-25  &49 &36 &15--1398   &0\farcs61  &9\farcs0   &J1743-1658 &J1924-2914\\
& &2019-08-19  &47 &37 &41--3638   &0\farcs24  &3\farcs9   &J1755-2232 &J1924-2914\\
& &2019-08-22  &46 &37 &41--3638   &0\farcs23  &3\farcs6   &J1755-2232 &J1924-2914\\
& &2019-09-06  &46 &37 &41--3638   &0\farcs23  &3\farcs6   &J1755-2232 &J1924-2914\\
\enddata
\end{deluxetable*}

\begin{deluxetable*}{llcccccccc}
\tablecaption{Details of Band 6 observations\label{tab:obs-b6}}
\tablehead{
\colhead{Setting}    &\colhead{Target}  &\colhead{Date}   & \colhead{\# Ant.}  &\colhead{Int.}    & \colhead{Baselines}	 & \colhead{Res.}& \colhead{Max Scale}& \colhead{Phase Cal.}& \colhead{Flux Cal.}\\
 \colhead{} & \colhead{}    &\colhead{}    & \colhead{} &
\colhead{[min]}    &\colhead{[m]} &\colhead{}&\colhead{}& \colhead{}& \colhead{} 
}
\startdata
B6-1  &IM Lup  &2018-11-28  &44 &37 &15--1241   &0\farcs36  &4\farcs8   &J1610-3958 &J1427-4206\\
&&2019-08-12  &42 &44 &41--3638   &0\farcs10  &1\farcs6   &J1610-3958 &	J1517-2422\\
&&2019-08-12  &43 &44 &41--3638   &0\farcs10  &1\farcs6   &J1610-3958 &	J1517-2422\\
&&2019-08-14  &43 &44 &41--3638   &0\farcs10  &1\farcs7   &J1610-3958 &	J1517-2422\\
\hline
&GM Aur/MWC 480  &2018-10-30  &48 &40 &15--1398   &0\farcs26  &3\farcs8   &J0438+3004 &J0510+1800\\
&&2018-11-17  &48 &40 &15--1398   &0\farcs26  &3\farcs8   &J0438+3004 &J0510+1800\\
&&2018-11-23  &49 &40 &15--1398   &0\farcs26  &3\farcs8   &J0438+3004 &J0510+1800\\
&&2019-08-26  &49 &47 &41--3638   &0\farcs10  &1\farcs6   &J0438+3004 &J0510+1800\\
&&2019-08-26  &49 &48 &41--3638   &0\farcs10  &1\farcs6   &J0438+3004 &J0510+1800\\
&&2019-08-27  &50 &48 &41--3638   &0\farcs10  &1\farcs6   &J0439+3045 &J0510+1800\\
&&2019-08-28  &50 &48 &38--3638   &0\farcs10  &1\farcs6   &J0439+3045 &J0510+1800\\
&&2019-08-28  &50 &48 &38--3638   &0\farcs10  &1\farcs6   &J0439+3045 &J0510+1800\\
\hline
&AS 209  &2018-12-02  &44 &41 &15--784   &0\farcs43  &5\farcs3   &J1733-1304 &J1751+0939\\
&&2019-08-15  &46 &43 &41--3638   &0\farcs10  &1\farcs7   &J1653-1551 &J1427-4206\\
&&2019-08-16  &46 &43 &41--3638   &0\farcs10  &1\farcs7   &J1653-1551 &J1517-2422\\
&&2019-08-16  &45 &43 &41--3638   &0\farcs10  &1\farcs6   &J1653-1551 &J1517-2422\\
\hline
&HD 163296  &2018-12-04  &43 &41 &15--784   &0\farcs44  &5\farcs9   &J1733-1304 &J1924-2914\\
&&2019-08-16  &45 &43 &41--3638   &0\farcs10  &1\farcs6   &J1755-2232 &J1924-2914\\
&&2019-08-16  &47 &43 &41--3638   &0\farcs10  &1\farcs6   &J1755-2232 &J1924-2914\\
&&2019-08-17  &47 &43 &41--3638   &0\farcs10  &1\farcs6   &J1755-2232 &J1924-2914\\
\hline\hline
B6-2  &IM Lup  &2019-04-07 &43 &16 &15--500   &0\farcs57  &6\farcs1   &J1610-3958 &J1517-2422\\
&&2019-04-09 &43 &36 &15--500   &0\farcs56  &6\farcs0   &J1610-3958 &	J1924-2914\\
&&2019-08-21  &44 &47 &41--3396  &0\farcs09  &1\farcs4   &J1733-1304 &J1924-2914\\
&&2019-08-25  &45 &47 &41--3396  &0\farcs09  &1\farcs4   &J1755-2232 &J1924-2914\\
&&2019-08-26  &45 &47 &41--3638  &0\farcs08  &1\farcs4   &J1755-2232 &J1924-2914\\
\hline
&GM Aur/MWC 480  &2018-10-31  &47 &47 &15--1398   &0\farcs22  &3\farcs4   &J0438+3004 &J0510+1800\\
&&2018-10-31  &47 &47 &15--1398   &0\farcs22  &3\farcs4   &J0438+3004 &J0510+1800\\
&&2019-08-08  &42 &47 &41--5894   &0\farcs08  &1\farcs2   &J0438+3004 &J0510+1800\\
&&2019-08-15  &48 &47 &41--3638   &0\farcs09  &1\farcs4   &J0438+3004 &J0510+1800\\
&&2019-08-15  &48 &47 &41--3638   &0\farcs09  &1\farcs4   &J0438+3004 &J0510+1800\\
&&2019-08-18  &43 &47 &41--3638   &0\farcs08  &1\farcs3   &J0438+3004 &J0510+1800\\
&&2019-08-20  &49 &47 &41--3638   &0\farcs08  &1\farcs2   &J0438+3004 &J0510+1800\\
&&2019-08-21  &43 &47 &41--3189   &0\farcs09  &1\farcs4   &J0438+3004 &J0510+1800\\
\hline
&AS 209  &2018-12-25  &45 &35 &15--500   &0\farcs57  &6\farcs7   &J1733-1304 &J1517-2422\\
&&2019-08-21  &45 &47 &41--3396   &0\farcs09  &1\farcs3   &J1733-1304 &J1517-2422\\
&&2019-08-21  &45 &47 &41--3396   &0\farcs09  &1\farcs4   &J1733-1304 &J1517-2422\\
\hline
&HD 163296  &2019-04-07  &43 &35 &15--500   &0\farcs57  &6\farcs1   &J1733-1304 &J1924-2914\\
&&2019-08-21  &44 &47 &41--3396   &0\farcs09  &1\farcs4   &J1733-1304 &J1924-2914\\
&&2019-08-25  &45 &47 &41--3396   &0\farcs09  &1\farcs4   &J1755-2232 &J1924-2914\\
&&2019-08-26  &45 &47 &41--3638   &0\farcs08  &1\farcs4   &J1755-2232 &J1924-2914\\
\enddata
\end{deluxetable*}

\section{Noise and noise correction}
\label{app:rms}

Table \ref{tab:wide_table_rms} reports the so called JvM \citep{Jorsater95} corrections ($\epsilon$) of the image residuals, and the measured RMS per beam and channel of JvM-corrected image cubes for each source and targeted molecular line. The JvM correction is equal to the ratio of the CLEAN beam and dirty beam effective areas, and it is needed because the dirty beam was highly non-Gaussian when combining the short and long baseline configurations. See \citet{Czekala21} for details of how ($\epsilon$) was derived. In cases where the hyperfine components were observed in the same SPW, we only report a single RMS and ($\epsilon$), since neither should vary substantially across these narrow SPWs.

\movetabledown=5cm
\begin{rotatetable*}
\begin{deluxetable*}{llc cc cc cc cc cc}
\tabletypesize{\scriptsize}
\tablecaption{MAPS sample RMS and JvM $\epsilon$}
\label{tab:wide_table_rms}
\tablehead{
  & &  & \multicolumn{2}{c}{IM Lup}& \multicolumn{2}{c}{GM Aur}& \multicolumn{2}{c}{AS 209}& \multicolumn{2}{c}{HD 163296}& \multicolumn{2}{c}{MWC 480}\\ 
\colhead{Set-up} & \colhead{Molecule} & \colhead{Line(s)} & \colhead{RMS} & \colhead{$\epsilon$} & \colhead{RMS} & \colhead{$\epsilon$} & \colhead{RMS} & \colhead{$\epsilon$} & \colhead{RMS} & \colhead{$\epsilon$} & \colhead{RMS} & \colhead{$\epsilon$}\\ 
 & &  & \colhead{[mJy beam$^{-1}$]} &  & \colhead{[mJy beam$^{-1}$]} &  & \colhead{[mJy beam$^{-1}$]} &  & \colhead{[mJy beam$^{-1}$]} &  & \colhead{[mJy beam$^{-1}$]} &  
}
\startdata 
B3-1& HC$^{15}$N& J=1--0 & 0.815 & 0.738& 1.718 & 1.005& 0.986 & 0.909& 0.851 & 0.950& 1.732 & 1.008\\ 
& H$^{13}$CN$^\dagger$& J=1--0 & 0.769 & 0.734& 1.631\tablenotemark{a} & 1.004& 0.915 & 0.901& 0.798 & 0.947& 1.848\tablenotemark{b} & 1.000\\ 
& H$^{13}$CO$^{+}$& J=1--0 & 0.775 & 0.729& 1.685 & 1.021& 0.953 & 0.903& 0.813 & 0.932& 1.816 & 1.000\\ 
& C$_2$H& N=1--0,J=$\frac{3}{2}$--$\frac{1}{2}$,F=2--1& 0.746 & 0.725& 1.522\tablenotemark{c} & 1.017& 0.904 & 0.893& 0.795 & 0.944& 1.700\tablenotemark{d} & 0.990\\ 
& & N=1--0,J=$\frac{3}{2}$--$\frac{1}{2}$,F=1--0& 0.737 & 0.721& 1.502\tablenotemark{e} & 1.016& 0.908 & 0.893& 0.799 & 0.945& 1.719\tablenotemark{b} & 0.996\\ 
& HCN$^\dagger$& J=1--0, F=1--1; F=2--1, F=0--1& 0.872 & 0.722& 2.052\tablenotemark{d} & 1.000& 0.919 & 0.842& 0.888 & 0.934& 1.944\tablenotemark{d} & 0.992\\ 
& HCO$^{+}$& J=1--0 & 0.859 & 0.723& 2.051 & 1.019& 0.880 & 0.810& 0.870 & 0.925& 1.907 & 0.979\\ 
& HC$_3$N& J=11--10 & 0.698 & 0.650& 1.765 & 1.062& 0.662 & 0.678& 0.643 & 0.732& 1.742 & 1.052\\ 
& H$_2$CO& (J$_{\rm{K}_{\rm{a}}, {\rm{K}_{\rm{c}}}}$)=6$_{15}$--6$_{16}$ & 0.549 & 0.526& 1.590 & 1.006& 0.696 & 0.689& 0.668 & 0.732& 1.580 & 0.990\\ 
\hline 
B3-2& CS& J=2--1& 0.640 & 0.639& 1.258 & 1.036& 0.646 & 0.671& 0.546 & 0.582& 1.266 & 1.039\\ 
& C$^{18}$O& J=1--0 & 0.565 & 0.561& 1.146 & 0.922& 0.670 & 0.642& 0.469 & 0.505& 1.129 & 0.920\\ 
& ${}^{13}$CO& J=1--0 & 0.552 & 0.560& 1.115 & 0.927& 0.471 & 0.511& 0.457 & 0.502& 1.113 & 0.926\\ 
& CH$_3$CN& J=6--5, K=0--2& 0.538 & 0.560& 1.081 & 0.922& 0.457 & 0.510& 0.438 & 0.509& 1.086 & 0.929\\ 
& & J=6--5, K=3& 0.553 & 0.560& 1.091 & 0.921& 0.460 & 0.510& 0.441 & 0.509& 1.094 & 0.928\\ 
& & J=6--5, K=4& 0.543 & 0.560& 1.076 & 0.920& 0.458 & 0.510& 0.443 & 0.510& 1.081 & 0.925\\ 
& & J=6--5, K=5& 0.541 & 0.560& 1.078 & 0.922& 0.457 & 0.510& 0.438 & 0.509& 1.081 & 0.924\\ 
& C$^{17}$O& J=1--0, F=$\frac{3}{2}$--$\frac{5}{2}$; F=$\frac{7}{2}$--$\frac{5}{2}$; F=$\frac{5}{2}$--$\frac{5}{2}$& 0.683 & 0.563& 1.363 & 0.896& 0.613 & 0.539& 0.561 & 0.508& 1.404 & 0.917\\ 
& CN& N=1--0, J=$\frac{3}{2}$--$\frac{1}{2}$, F=$\frac{3}{2}$--$\frac{1}{2}$; F=$\frac{5}{2}$--$\frac{3}{2}$& 0.827 & 0.573& 1.251 & 0.751& 0.733 & 0.538& 0.647 & 0.488& 1.227\tablenotemark{d} & 0.741\\ 
& & N=1--0, J=$\frac{3}{2}$--$\frac{1}{2}$, F=$\frac{1}{2}$--$\frac{1}{2}$ & 0.806 & 0.573& 1.233 & 0.752& 0.688 & 0.513& 0.640 & 0.489& 1.215\tablenotemark{d} & 0.741\\ 
& & N=1--0, J=$\frac{3}{2}$--$\frac{1}{2}$, F=$\frac{3}{2}$--$\frac{3}{2}$ & 0.831 & 0.575& 1.260 & 0.752& 0.700 & 0.513& 0.676 & 0.514& 1.233\tablenotemark{d} & 0.741\\ 
\hline 
B6-1& DCN$^\dagger$ & J=3--2 & 0.886 & 0.621& 1.064 & 0.872& 0.391 & 0.360& 0.511 & 0.459& 1.081 & 0.856\\ 
& ${}^{13}$CN$^*$& N=2--1, J=$\frac{3}{2}$--$\frac{1}{2}$& 0.876 & 0.619& 1.063 & 0.874& 0.437 & 0.448& 0.508 & 0.459& 1.074 & 0.854\\ 
& H$_2$CO& (J$_{\rm{K}_{\rm{a}}, {\rm{K}_{\rm{c}}}}$)=3$_{03}$--2$_{02}$ & 0.816 & 0.627& 0.952 & 0.863& 0.344 & 0.346& 0.458 & 0.455& 0.989 & 0.865\\ 
& C$^{18}$O& J=2--1& 0.732 & 0.600& 0.878 & 0.809& 0.339 & 0.347& 0.415 & 0.436& 0.914 & 0.823\\ 
& $^{13}$CO& J=2--1 & 0.977 & 0.575& 1.210 & 0.804& 0.471 & 0.347& 0.581 & 0.436& 1.261 & 0.819\\ 
& CH$_3$CN& J=12--11, K=0--2 & 1.052 & 0.626& 1.259 & 0.866& 0.522 & 0.446& 0.603 & 0.460& 1.311 & 0.877\\ 
& & J=12--11, K=3& 1.057 & 0.625& 1.276 & 0.876& 0.529 & 0.444& 0.578 & 0.435& 1.349 & 0.901\\ 
& CO& J=2--1 & 1.051 & 0.594& 1.318 & 0.800& 0.562 & 0.363& 0.639 & 0.445& 1.305 & 0.798\\ 
& N$_2$D+$^\dagger$& J=3--2 & 1.265 & 0.605& 1.225 & 0.738& 0.576 & 0.343& 0.665 & 0.436& 1.210 & 0.729\\ 
\hline 
B6-2& $c$-C$_3$H$_2$& (J$_{\rm{K}_{\rm{a}}, {\rm{K}_{\rm{c}}}}$)=7$_{07}$--6$_{16}$ & 0.303 & 0.269& 0.649 & 0.559& 0.390 & 0.296& 0.376 & 0.339& 0.893 & 0.697\\ 
& & (J$_{K+},_{K-}$) 6$_{15}$--5$_{24}$& 0.308 & 0.268& 0.655 & 0.560& 0.397 & 0.295& 0.383 & 0.340& 0.909 & 0.697\\ 
& & (J$_{K+},_{K-}$) 6$_{25}$--5$_{14}$& 0.308 & 0.268& 0.659 & 0.560& 0.396 & 0.294& 0.385 & 0.340& 0.908 & 0.696\\ 
& C$_2$H& N=3--2, J=$\frac{5}{2}$--$\frac{3}{2}$, F=3--2; F=2--1; F=2--2 & 0.398 & 0.247& 0.951 & 0.556& 0.523 & 0.278& 0.503 & 0.332& 0.924 & 0.545\\ 
& & N=3--2, J=$\frac{7}{2}$--$\frac{5}{2}$,F=4--3; F=3--2& 0.407 & 0.247& 0.974 & 0.556& 0.534 & 0.278& 0.509 & 0.331& 0.946 & 0.545\\ 
& HC$_3$N& J=29--28 & 0.291 & 0.241& 0.740 & 0.557& 0.399 & 0.277& 0.326 & 0.282& 0.747 & 0.552\\ 
& HCN$^\dagger$& J=3--2, F=3--2; F=3--3; F=2--2& 0.342 & 0.248& 0.797 & 0.573& 0.461 & 0.281& 0.398 & 0.303& 0.757 & 0.535\\ 
\hline 
\enddata 
\tablecomments{Beams are 0\farcs3$\times$ 0\farcs3 for Band 3 and 0\farcs15$\times$ 0\farcs15 Band 6 observations except those with those noted which have beams of: $^a${0\farcs33 $\times$ 0\farcs33}, $^b${0\farcs315 $\times$ 0\farcs314}, $^c${0\farcs33 $\times$ 0\farcs326}, $^d${0\farcs315 $\times$ 0\farcs315}, $^e${0\farcs33 $\times$ 0\farcs315}.} 
\tablenotetext{}{$\dagger$ See \citet{Cataldi21} for the full list of hyperfine components.}
\tablenotetext{}{$*$ See Table \ref{tab:lines-b6} for a full list of observed $^{13}$CN hyperfine components.}
\end{deluxetable*}
\end{rotatetable*}

\section{Disk-integrated line fluxes}
\label{app:lines}

This appendix reports the disk integrated line fluxes or 3$\sigma$ upper limits as well as the peak SNR. The integrated fluxes were generated from the 0\farcs3 image cubes using the relevant \clean-masks for each disk. For weaker lines we only used the  \clean-mask channels were C$^{18}$O showed strong emission, $>$5$\sigma$.

In the case of hyperfine components, the SNR and fluxes sre reported for each component if they are readily separable towards all disks. In most cases there is some line overlap in at least on disk, however, and in these cases the integrated flux includes all listed hyperfine components, and the peak SNR refers to the strongest line component in the cube. The listed uncertainties do not take into account calibration uncertainties of $\sim$10\%. 

Note that because these line fluxes were generated using an automated pipeline, they should be viewed as approximations. The same pipeline also generates very conservative noise estimates compared to a detailed analysis. As a result, several lines that are reported as upper limits here are revealed to be low SNR detections upon closer inspection and/or using tools such as matched filter analysis \citep{Loomis18a}. For precise line fluxes, upper limits, and radial profiles, the reader is advised to consult \citet{Zhang21} for CO isotopologues, \citet{Guzman21} for HCN, C$_2$H, and H$_2$CO lines, \citet{Cataldi21} for DCN and N$_2$D$^+$, \citet{Ilee21} for HC$_3$N, CH$_3$N, and $c$-C$_3$H$_2$, \citet{Bergner21} for CN, \citet{LeGal21} for CS (as well as upper limits on other S-bearing molecules), and \citet{Aikawa21} for HCO$^+$ and H$^{13}$CO$^+$.

\movetabledown=5cm
\begin{rotatetable*}
\begin{deluxetable*}{llc cc cc cc cc cc}
\tabletypesize{\scriptsize}
\tablecaption{{Approximate disk-integrated line fluxes and upper limits extracted using an automated procedure (see text)}}
\label{tab:wide_table_fluxes}
\tablehead{
  & &  & \multicolumn{2}{c}{IM Lup}& \multicolumn{2}{c}{GM Aur}& \multicolumn{2}{c}{AS 209}& \multicolumn{2}{c}{HD 163296}& \multicolumn{2}{c}{MWC 480}\\ 
\colhead{Set-up} & \colhead{Molecule} & \colhead{Line(s)} & \colhead{Peak} & \colhead{Flux} & \colhead{Peak} & \colhead{Flux} & \colhead{Peak} & \colhead{Flux} & \colhead{Peak} & \colhead{Flux} & \colhead{Peak} & \colhead{Flux}\\ 
 & & & \colhead{[SNR]}& \colhead{[Jy km s$^{-1}$]}& \colhead{[SNR]}& \colhead{[Jy km s$^{-1}$]}& \colhead{[SNR]}& \colhead{[Jy km s$^{-1}$]}& \colhead{[SNR]}& \colhead{[Jy km s$^{-1}$]}& \colhead{[SNR]}& \colhead{[Jy km s$^{-1}$]} 
}
\startdata 
B3-1& HC$^{15}$N& J=1--0& 4.7&$<$0.029& 3.8&$<$0.022& 4.9&$<$0.016& 4.3&$<$0.031& 3.8&$<$0.022\\ 
& H$^{13}$CN& J=1--0& 4.4&$<$0.045& 5.4&$<$0.035& 3.8&$<$0.025& 4.7&$<$0.048& 5.7&$<$0.035\\ 
& H$^{13}$CO$^{+}$& J=1--0& 4.8&$<$0.027$^*$& 4.7& 0.026$\pm$0.011& 5.2& 0.020$\pm$0.008& 3.7&$<$0.030$^*$& 3.9&$<$0.022$^*$\\ 
& C$_2$H& N=1--0,J=$\frac{3}{2}$--$\frac{1}{2}$,F=2--1& 3.8& 0.051$\pm$0.013& 5.7& 0.047$\pm$0.021$^\dagger$& 17.2& 0.070$\pm$0.008& 11.4& 0.107$\pm$0.015& 13.9& 0.051$\pm$0.010\\ 
& & N=1--0,J=$\frac{3}{2}$--$\frac{1}{2}$,F=1--0& 3.7&$<$0.027$^*$& 3.9&$<$0.021$^*$& 11.5& 0.050$\pm$0.008& 6.6& 0.062$\pm$0.015& 9.1&$<$0.021$^*$\\ 
& HCN& J=1--0, F=1--1; F=2--1, F=0--1& 8.4& 0.349$\pm$0.027& 7.0& 0.207$\pm$0.021& 18.3& 0.242$\pm$0.015& 18.2& 0.713$\pm$0.027& 16.9& 0.247$\pm$0.022\\ 
& HCO$^{+}$& J=1--0& 33.4& 0.595$\pm$0.016& 27.0& 0.580$\pm$0.012& 30.1& 0.198$\pm$0.009& 33.4& 1.092$\pm$0.016& 26.5& 0.452$\pm$0.013\\ 
& HC$_3$N& J=11--10& 3.8&$<$0.035& 7.8& 0.047$\pm$0.013& 18.8& 0.123$\pm$0.010& 18.4& 0.144$\pm$0.019& 16.6& 0.083$\pm$0.013\\ 
& H$_2$CO& (J$_{\rm{K}_{\rm{a}}, {\rm{K}_{\rm{c}}}}$)=6$_{15}$--5$_{16}$& 3.7&$<$0.038& 3.8&$<$0.028& 3.3&$<$0.021& 3.9&$<$0.041& 3.4&$<$0.028\\ 
\hline 
B3-2& CS& J=2--1& 12.2& 0.341$\pm$0.016& 20.9& 0.271$\pm$0.011& 24.6& 0.165$\pm$0.009& 20.5& 0.307$\pm$0.017& 10.7& 0.056$\pm$0.011\\ 
& C$^{18}$O& J=1--0& 9.6& 0.161$\pm$0.019& 18.6& 0.149$\pm$0.013& 5.8& 0.073$\pm$0.010& 41.5& 0.976$\pm$0.020& 36.2& 0.410$\pm$0.013\\ 
& ${}^{13}$CO& J=1--0& 41.7& 1.337$\pm$0.018& 48.3& 0.836$\pm$0.013& 30.2& 0.376$\pm$0.010& 78.1& 3.573$\pm$0.020& 73.3& 1.522$\pm$0.013\\ 
& CH$_3$CN& J=6--5, K=0--2& 3.8&$<$0.063& 3.8&$<$0.043$^*$& 4.4&$<$0.032$^*$& 6.0&$<$0.063$^*$& 4.7&$<$0.044$^*$\\ 
& & J=6--5, K=3& 3.7&$<$0.037& 3.9&$<$0.025& 3.4&$<$0.019& 4.9&$<$0.037& 4.0&$<$0.026\\ 
& & J=6--5, K=4& 5.0&$<$0.036& 3.8&$<$0.025& 4.3&$<$0.019& 3.9&$<$0.038& 3.4&$<$0.026\\ 
& & J=6--5, K=5& 3.7&$<$0.036& 3.7&$<$0.025& 3.8&$<$0.019& 3.5&$<$0.037& 3.7&$<$0.026\\ 
& C$^{17}$O& J=1--0, F=$\frac{3}{2}$--$\frac{5}{2}$; F=$\frac{7}{2}$--$\frac{5}{2}$; F=$\frac{5}{2}$--$\frac{5}{2}$& 5.1&$<$0.071$^*$& 7.6& 0.083$\pm$0.024& 4.6&$<$0.037$^*$& 27.2& 0.675$\pm$0.036& 20.7& 0.296$\pm$0.025\\ 
& CN& N=1--0, J=$\frac{3}{2}$--$\frac{1}{2}$, F=$\frac{3}{2}$--$\frac{1}{2}$; F=$\frac{5}{2}$--$\frac{3}{2}$& 8.8& 0.581$\pm$0.038& 5.5& 0.215$\pm$0.029& 19.3& 0.406$\pm$0.022& 15.1& 1.585$\pm$0.043& 21.0& 0.479$\pm$0.029\\ 
& & N=1--0, J=$\frac{3}{2}$--$\frac{1}{2}$,  F=$\frac{1}{2}$--$\frac{1}{2}$& 6.7& 0.097$\pm$0.027& 6.0& 0.064$\pm$0.020& 10.8& 0.102$\pm$0.015& 7.7& 0.343$\pm$0.031& 9.6& 0.114$\pm$0.021\\ 
& & N=1--0, J=$\frac{3}{2}$--$\frac{1}{2}$,  F=$\frac{3}{2}$--$\frac{3}{2}$& 5.0& 0.106$\pm$0.027& 3.5&$<$0.041& 10.7& 0.131$\pm$0.016& 6.5& 0.318$\pm$0.031& 8.2& 0.121$\pm$0.021\\ 
\hline 
B6-1& DCN& J=3--2& 5.2& 0.446$\pm$0.125$^\dagger$& 5.2& 0.279$\pm$0.075$^\dagger$& 21.8& 0.278$\pm$0.030& 7.6& 0.751$\pm$0.131$^\dagger$& 10.1& 0.436$\pm$0.090$^\dagger$\\ 
& ${}^{13}$CN& N=2--1, J=$\frac{3}{2}$--$\frac{1}{2}$& 4.8&$<$0.271& 4.5&$<$0.151& 4.9&$<$0.128& 9.2&$<$0.227& 5.2&$<$0.151\\ 
& H$_2$CO& (J$_{\rm{K}_{\rm{a}}, {\rm{K}_{\rm{c}}}}$)=3$_{03}$--2$_{02}$& 14.4& 0.852$\pm$0.061& 29.9& 0.913$\pm$0.039& 16.0& 0.299$\pm$0.027& 11.2& 0.902$\pm$0.062& 6.2& 0.151$\pm$0.041\\ 
& C$^{18}$O& J=2--1& 32.8& 1.592$\pm$0.060& 52.0& 1.092$\pm$0.039& 34.0& 0.538$\pm$0.027& 85.8& 5.783$\pm$0.051& 87.0& 3.017$\pm$0.041\\ 
& $^{13}$CO& J=2--1& 54.3& 8.370$\pm$0.083& 73.4& 5.028$\pm$0.048& 58.6& 2.269$\pm$0.038& 84.9& 15.885$\pm$0.080& 103.7& 8.361$\pm$0.057\\ 
& CH$_3$CN& J=12--11, K=0--2& 5.1&$<$0.279$^*$& 5.4&$<$0.177$^*$& 5.3&$<$0.136$^*$& 5.8&$<$0.271$^*$& 5.2&$<$0.189$^*$\\ 
& & J=12--11, K=3& 4.9&$<$0.164& 5.3&$<$0.105$^*$& 5.0&$<$0.081& 4.2&$<$0.162$^*$& 4.9& 0.234$\pm$0.109$^\dagger$\\ 
& CO& J=2--1& 89.7& 22.342$\pm$0.094& 99.9& 19.844$\pm$0.074& 118.4& 7.790$\pm$0.045& 142.5& 45.246$\pm$0.102& 217.6& 23.226$\pm$0.063\\ 
& N$_2$D+& J=3--2& 4.6& 0.559$\pm$0.187$^\dagger$& 3.9&$<$0.137$^*$& 5.3& 0.450$\pm$0.083$^\dagger$& 5.9& 0.637$\pm$0.173$^\dagger$& 4.4&$<$0.138$^*$\\ 
\hline 
B6-2& $c$-C$_3$H$_2$& (J$_{\rm{K}_{\rm{a}}, {\rm{K}_{\rm{c}}}}$)=7$_{07}$--6$_{16}$& 5.5&$<$0.076$^*$& 5.5&$<$0.110$^*$& 27.1& 0.268$\pm$0.032& 15.7& 0.227$\pm$0.054& 15.1& 0.489$\pm$0.110$^\dagger$\\ 
& & (J$_{\rm{K}_{\rm{a}}, {\rm{K}_{\rm{c}}}}$)=6$_{15}$--5$_{24}$& 6.2&$<$0.077& 4.0&$<$0.111$^*$& 6.9& 0.267$\pm$0.058$^\dagger$& 6.1& 0.250$\pm$0.093$^\dagger$& 5.3&$<$0.115\\ 
& & (J$_{\rm{K}_{\rm{a}}, {\rm{K}_{\rm{c}}}}$)=6$_{25}$--5$_{14}$& 5.1&$<$0.077& 4.8&$<$0.112$^*$& 15.1& 0.162$\pm$0.032& 8.7& 0.745$\pm$0.106$^\dagger$& 10.0& 0.464$\pm$0.113$^\dagger$\\ 
& C$_2$H& N=3--2, J=$\frac{5}{2}$--$\frac{3}{2}$, F=3--2; F=2--1; F=2--2& 9.5& 0.473$\pm$0.088& 8.9& 0.409$\pm$0.133& 55.8& 1.904$\pm$0.075& 48.6& 2.852$\pm$0.125& 39.3& 1.210$\pm$0.133\\ 
& & N=3--2, J=$\frac{7}{2}$--$\frac{5}{2}$,F=4--3; F=3--2& 14.0& 0.717$\pm$0.072& 12.9& 0.573$\pm$0.110& 60.6& 2.166$\pm$0.060& 57.8& 3.467$\pm$0.100& 42.7& 1.478$\pm$0.110\\ 
& HC$_3$N& J=29--28& 4.4&$<$0.077& 8.4& 0.395$\pm$0.121$^\dagger$& 6.6& 0.256$\pm$0.060$^\dagger$& 24.6& 0.173$\pm$0.052& 9.8& 0.129$\pm$0.064\\ 
& HCN& J=3--2, F=3--2; F=3--3; F=2--2& 49.8& 2.443$\pm$0.073& 48.6& 1.835$\pm$0.107& 85.0& 3.010$\pm$0.061& 124.5& 7.472$\pm$0.098& 68.8& 2.561$\pm$0.110\\ 
\hline 
\enddata 
\tablenotetext{}{$\dagger$ Only channels with at least 5$\sigma$ flux in C$^{18}$O were included to calculate these fluxes to reduce noise.}
\tablenotetext{}{$^*$ These lines are detected when the flux extraction is optimized. See \citep{Zhang21,Guzman21,Ilee21,Cataldi21,Bergner21,LeGal21, Aikawa21} for the precise line fluxes.}
\end{deluxetable*}
\end{rotatetable*}

\bibliography{mybib}{}



\end{document}